# Approximation Algorithms and Hardness of the $k$-Route Cut Problem[*]

Julia Chuzhoy[†]    Yury Makarychev[‡]    Aravindan Vijayaraghavan[§]    Yuan Zhou[¶]

November 6, 2018

**Abstract**

We study the $k$-route cut problem: given an undirected edge-weighted graph $G = (V, E)$, a collection $\{(s_1, t_1), (s_2, t_2), \ldots, (s_r, t_r)\}$ of source-sink pairs, and an integer connectivity requirement $k$, the goal is to find a minimum-weight subset $E'$ of edges to remove, such that the connectivity of every pair $(s_i, t_i)$ falls below $k$. Specifically, in the edge-connectivity version, EC-kRC, the requirement is that there are at most $(k-1)$ edge-disjoint paths connecting $s_i$ to $t_i$ in $G \setminus E'$, while in the vertex-connectivity version, VC-kRC, the same requirement is for vertex-disjoint paths. Prior to our work, poly-logarithmic approximation algorithms have been known for the special case where $k \leq 3$, but no non-trivial approximation algorithms were known for any value $k > 3$, except in the single-source setting. We show an $O(k \log^{3/2} r)$-approximation algorithm for EC-kRC with uniform edge weights, and several polylogarithmic bi-criteria approximation algorithms for EC-kRC and VC-kRC, where the connectivity requirement $k$ is violated by a constant factor. We complement these upper bounds by proving that VC-kRC is hard to approximate to within a factor of $k^\epsilon$ for some fixed $\epsilon > 0$.

We then turn to study a simpler version of VC-kRC, where only one source-sink pair is present. We give a simple bi-criteria approximation algorithm for this case, and show evidence that even this restricted version of the problem may be hard to approximate. For example, we prove that the single source-sink pair version of VC-kRC has no constant-factor approximation, assuming Feige's Random $\kappa$-AND assumption.

## 1 Introduction

Multi-commodity flows and cuts are among the most extensively studied graph optimization problems. Due to their rich connections to many combinatorial optimization problems, algorithms for various versions of flow and cut problems are a powerful and a widely used toolkit. One of the central problems in this area is *minimum multicut*: given an $n$-vertex graph $G = (V, E)$ with non-negative weights $w_e$ on edges $e \in E$ and a collection $\{(s_1, t_1), (s_2, t_2), \ldots, (s_r, t_r)\}$ of source-sink pairs, find a minimum-weight subset $E'$ of edges to delete, so that each pair $(s_i, t_i)$ is disconnected in the resulting graph $G \setminus E'$. The dual to minimum multicut is the *maximum multi-commodity flow* problem, where the goal is to find a maximum flow between the pairs $(s_i, t_i)$, with the restriction that each edge $e$ carries at most $w_e$ flow units. It is easy to see that minimum multicut can be viewed as revealing a bottleneck in the routing capacity of $G$, as the value of any multi-commodity flow cannot exceed the value of the minimum multicut in $G$. A fundamental result, due to Leighton and Rao [LR99] and Garg, Vazirani and Yannakakis [GVY95] shows that the value of minimum multicut is within an $O(\log r)$ factor of that of maximum multicommodity flow in any graph, where $r$ is the number of the source-sink pairs. This result can be seen as an extension of the famous min-cut max-flow theorem to the multicommodity setting, and it also gives an efficient $O(\log r)$-approximation algorithm for minimum multicut — the best currently known approximation guarantee for it.

---

[*]Extended abstract is to appear in SODA 2012

[†]Toyota Technological Institute, Chicago, IL 60637. Email: cjulia@ttic.edu. Supported in part by NSF CAREER award CCF-0844872 and Sloan Research Fellowship.

[‡]Toyota Technological Institute, Chicago, IL 60637. Email: yury@ttic.edu.

[§]Department of Computer Science, Princeton University. Email: aravindv@cs.princeton.edu. Work done while visiting Toyota Technological Institute, Chicago

[¶]Computer Science Department, Carnegie Mellon University, Pittsburgh, PA. Email: yuanzhou@cs.cmu.edu. Work done while visiting Toyota Technological Institute, Chicago

In this paper we study a natural generalization of minimum multicut - the *minimum $k$-route cut* problem. In this problem, the input again consists of an $n$-vertex graph $G = (V, E)$ with non-negative weights $w_e$ on edges $e \in E$, and a collection $\{(s_1, t_1), (s_2, t_2), \ldots, (s_r, t_r)\}$ of $r$ source-sink pairs. Additionally, we are given an integral connectivity threshold $k > 0$. The goal is to find a minimum-weight subset $E' \subseteq E$ of edges to delete, such that the connectivity of each pair $(s_i, t_i)$ falls below $k$ in the resulting graph $G \setminus E'$. We study two versions of this problem: in the edge-connectivity version (EC-kRC), the requirement is that for each $1 \leq i \leq r$, the number of **edge-disjoint** paths connecting $s_i$ to $t_i$ in graph $G \setminus E'$ is less than $k$. In the vertex-connectivity version (VC-kRC), the requirement is that the number of **vertex-disjoint** paths connecting $s_i$ to $t_i$ is less than $k$. It is not hard to see that VC-kRC captures EC-kRC as a special case (see Section A), and hence is more general. It is also easy to see that minimum multicut is a special case of both EC-kRC and VC-kRC, with the connectivity requirement $k = 1$. We also consider a special case of EC-kRC, where all edges have unit weight, and we refer to it as the *uniform* EC-kRC. We note that for VC-kRC, the uniform and the non-uniform edge-weight versions are equivalent up to a small loss in the approximation factor, as shown in Section B, and so we do not distinguish between them.

The primary motivation for studying $k$-route cuts comes from multi-commodity flows in fault tolerant settings, where the resilience to edge and node failures is important. An elementary $k$-route flow between a pair $s$ and $t$ of vertices is a set of $k$ disjoint paths connecting $s$ to $t$. A $k$-route (st)-flow is just a combination of such elementary $k$-route flows, where each elementary flow is assigned some fractional value. This is a natural generalization of the standard (st)–flows, which ensures that the flow is resilient to the failure of up to $(k - 1)$ edges or vertices. Multi-route flows were first introduced by Kishimoto [Kis96], and have since been studied in the context of communication networks [BCSK07, BCK03, ACKN07]. In a series of papers, Kishimoto [Kis96], Kishimoto and Takeuchi [KT93] and Aggarwal and Orlin [AO02] have developed a number of efficient algorithms for computing maximum multi-route flows. As in the case of standard flows, we can extend $k$-route (st)-flows to the multi-commodity setting, where the goal is to maximize the total $k$-route flow between all source-destination pairs. It is easy to see that the minimum $k$-route cut is a natural upper bound on the maximum $k$-route flow – just like minimum multicut upper-bounds the value of the maximum multi-commodity flow. Hence, as in the case with the standard multicut, multi-route cuts can be seen as revealing the network bottleneck, and so the minimum $k$-route cut in a graph captures the robustness of real-life computer and transportation networks.

The first approximation algorithm for the EC-kRC problem, due to Chekuri and Khanna [CK08], achieved a factor $O(\log^2 n \log r)$-approximation for the special case where $k = 2$, by rounding a Linear Programming relaxation. This was improved by Barman and Chawla [BC10] to give an $O(\log^2 r)$-approximation algorithm for the same version, by generalizing the region-growing LP-rounding scheme of [LR99, GVY95]. They note that it seems unlikely that their algorithm can be extended to handle higher values of $k$ using similar techniques. Very recently, Kolman and Scheideler [KS11] obtained a $O(\log^3 r)$ approximation to EC-3RC ($k = 3$ case) from the linear program of [BC10] by using a multi-level ball growing rounding. To the best of our knowledge, no approximation algorithms with sub-polynomial (in $n$) guarantees are known for any variant of the problem, for any value $k > 3$, except in the single-source setting that we discuss later. Our first result is an $O(k \log^{1.5} r)$-approximation algorithm for the uniform version of EC-kRC.

Since the problem appears to be computationally difficult, it is natural to turn to bi-criteria approximation, by slightly relaxing the connectivity requirement. Given parameters $\alpha, \beta > 1$, we say that an algorithm is an $(\alpha, \beta)$-bi-criteria approximation for EC-kRC (or VC-kRC), if it is guaranteed to produce a valid $k'$-route cut of weight at most $\beta \cdot \mathsf{OPT}$, where $k' \leq \alpha k$, and $\mathsf{OPT}$ is the value of the optimal $k$-route cut. Indeed, we can do much better in the bi-criteria setting: we obtain a $(1 + \delta, O(\frac{1}{\delta} \log^{1.5} r))$-bi-criteria approximation for any constant $0 < \delta < 1$, for the uniform EC-kRC problem (notice that the factors do not depend on $k$). When edge weights are arbitrary, we obtain a $\left(2, \tilde{O}(\log^{2.5} r)\right)$-bi-criteria approximation in $n^{O(k)}$ time, and an $\left(O(\log r), O(\log^3 r)\right)$-bi-criteria approximation in time polynomial in $n$ and $k$. We also show an $O(\log^{1.5} r)$-approximation for the special case where $k = 2$, thus slightly improving the result of [BC10]. The previously known upper bounds and our results for EC-kRC are summarized in Table 1.

|  | Previous results | Current paper |
| --- | --- | --- |
| $k = 2$ | $O(\log^2 r)$ [BC10] | $O(\log^{1.5} r)$ |
| $k = 3$ | $O(\log^3 r)$ [KS11] |  |
| arbitrary $k$, uniform | - | $O(k \log^{1.5} r)$, $\left(1 + \delta, O\left(\frac{1}{\delta} \log^{1.5} r\right)\right)$ for any constant $0 < \delta < 1$ |
| arbitrary $k$, general | - | $\left(2, O(\log^{2.5} r \log \log r)\right)$ in time $n^{O(k)}$; $\left(O(\log r), O(\log^3 r)\right)$ in poly$(n)$-time |

Table 1: Upper bounds for EC-kRC. Running time is polynomial in $n$ and $k$ unless stated otherwise.

We note that on the inapproximability side, it is easy to show that for any value of $k$, EC-kRC is at least as hard as minimum multicut, up to small constant factors[1]. While minimum multicut is known to be hard to approximate up to any constant factor assuming the Unique Games Conjecture [KV05, CKK$^+$06], it is only known to be NP-hard to approximate to within a small constant factor [DJP$^+$94]. In fact one of the motivations for studying $k$-route cuts is that inapproximability results may yield insights into approximation hardness of multicut.

We now turn to the more general VC-kRC problem. The $O(\log^2 n \log r)$-approximation of [CK08], and the $O(\log^2 r)$-approximation of [BC10] for 2-route cuts extend to the vertex-connectivity ver-sion as well, as does our $O(\log^{1.5} r)$-approximation algorithm. Prior to our work, no non-trivial approximation algorithms were known for any higher values of $k$. In this paper, we show a $\left(2, \tilde{O}(kd \log^{2.5} r)\right)$-bi-criteria approximation algorithm for VC-kRC, with running time $n^{O(k)}$, where $d$ is the maximum number of demand pairs in which any terminal participates. We note that, as in the case of EC-kRC, for any value of $k$, VC-kRC is at least as hard to approximate as minimum multicut (up to small constant factors), and to the best of our knowledge, no other inapproximability results have been known for this problem. We show that VC-kRC is hard to approximate up to any factor better than $\Omega(k^\epsilon)$, for some constant $\epsilon > 0$. Our results for VC-kRC are summarized in Table 2.

---

[1] A simple reduction replaces every vertex $v$ of the multicut instance by a set $S_v$ of $M$ vertices, where $M \gg k$, and every edge $(u, v)$ by a set of $M^2$ edges connecting every vertex of $S_v$ to every vertex of $S_u$.

|  | Previous results | Current paper |
| --- | --- | --- |
| $k=2$ | $O(\log^2 r)$ [BC10] | $O(\log^{1.5} r)$ |
| arbitrary $k$ | APX-hard [DJP$^+$94] <br> no constant factor approximation under UGC [KV05, CKK$^+$06] | $\left(2, O(dk \log^{2.5} r \log \log r)\right)$-approximation algorithm, running time $n^{O(k)}$, where $d$ is the maximum number of demand pairs in which any terminal participates <br> $\Omega(k^\epsilon)$-hardness for some constant $\epsilon > 0$ |

Table 2: Results for VC-kRC.

In order to better understand the multi-route cut problem computationally, it is instructive to consider a simpler special case, where we are only given a single source-sink pair $(s, t)$. We refer to this special case of VC-kRC and EC-kRC as (st)-VC-kRC and (st)-EC-kRC, respectively. As in the general case, it is easy to see that (st)-EC-kRC can be cast as a special case of (st)-VC-kRC. When the connectivity requirement $k$ is constant, both problems can be solved efficiently as follows: guess a set $E'$ of $(k-1)$ edges, and compute the minimum edge (st)–cut in graph $G \setminus E'$. The algorithm for (st)-VC-kRC is similar except that we guess a set $V'$ of $(k-1)$ vertices, and compute the minimum edge (st)–cut in graph $G \setminus V'$. However, for larger values of $k$, only a $2(k-1)$-approximation is known for (st)-EC-kRC, for the special case where the edge weights are uniform, due to Bruhn et al [BČH$^+$08][2]. Barman and Chawla [BC10] show that a generalization of (st)-EC-kRC where edges are allowed to have capacities is NP-hard. As no good approximation guarantees are known for the problem, it is natural to turn to bi-criteria approximation. For general values of $k$, Barman and Chawla [BC10] have shown a $(4,4)$-bi-criteria approximation algorithm for (st)-EC-kRC, and a $(2,2)$-bi-criteria approximation for uniform (st)-EC-kRC. In fact all these algorithms extend to a single-source multiple-sink scenario, except that the factor $(4, 4)$-approximation requires that the number of terminals is constant. In this paper we focus on the more general node-connectivity version of the problem. We start by showing a simple factor $(k+1)$-approximation algorithm for (st)-VC-kRC, and a factor $\left(1 + \frac{1}{c}, 1 + c\right)$-bi-criteria approximation for any constant $c$. We then complement these upper bounds by providing evidence that the problem is hard to approximate. Specifically, we show that for any constant $C$, there is no $(1 + \gamma, C)$-bi-criteria approximation for (st)-VC-kRC, assuming Feige's Random $\kappa$-AND Hypothesis, where $\gamma$ is some small constant depending on $C$. We also show that a factor $\rho$ approximation algorithm for (st)-VC-kRC would lead to a factor $2\rho^2$-approximation for the Densest $\kappa$-Subgraph problem. These inapproximability results are inspired by the recent work of Arora et al. [AAM$^+$11], who have ruled out a constant factor approximation for Densest $\kappa$-Subgraph assuming Feige's Random $\kappa$-AND hypothesis.

---

[2]This result also extends to the single-source multiple-sinks setting.

Recall that the Densest $\kappa$-Subgraph problem takes as input a graph $G(V, E)$ on $n$ vertices and a parameter $\kappa$, and asks for a subgraph of $G$ on at most $\kappa$ vertices containing the maximum number of edges. While it is a fundamental graph optimization problem, there is a huge gap between the best known approximation algorithms and the known inapproximability results. The current best approximation algorithm, due to Bhaskara et al. [BCC+10] achieves an $O(n^{1/4+\epsilon})$-approximation in time $n^{O(1/\epsilon)}$ for any constant $\epsilon > 0$. On the negative side, Feige [Fei02] showed a small constant factor inapproximability using the random 3-SAT assumption, and later Khot [Kho04] used quasi-random PCPs to rule out a PTAS, assuming $\mathsf{NP} \not\subseteq \bigcap_{\epsilon>0} \mathsf{BPTIME}(2^{n^\epsilon})$. Raghavendra and Steurer [RS10] and Alon et al. [AAM+11] ruled out constant factor approximation algorithms for Densest $\kappa$-Subgraph under other less standard complexity assumptions. The Densest $\kappa$-Subgraph problem can also be generalized to $\lambda$-uniform hypergraphs, where the goal is again to find a subset of $\kappa$ vertices containing maximum possible number of hyperedges. We show that for any constant $\lambda \geq 2$, a factor $\rho$ approximation algorithm for (st)-VC-kRC would lead to a factor $(2\rho^\lambda)$-approximation for the $\lambda$-uniform Densest $\kappa$-subgraph. We note that Applebaum [App11] has recently shown that for $\lambda \geq 3$, the $\lambda$-uniform Densest $\kappa$-subgraph problem is hard to approximate to within $n^\epsilon$-factor for some constant $\epsilon > 0$ assuming the existence of a certain family of one-way functions.

All our inapproximability results for (st)-VC-kRC are proved using a "proxy" problem, Small Set Vertex Expansion (SSVE). In this problem, we are given a bipartite graph $G = (U, V, E)$ and a parameter $0 \leq \alpha \leq 1$. The goal is to find a subset $S \subseteq U$ of $\alpha \cdot |U|$ vertices, while minimizing the number of its neighbors $|\Gamma(S)|$. We show an approximation-preserving reduction from SSVE to (st)-VC-kRC, and then prove inapproximability results for the SSVE problem. In particular, we show that approximating SSVE is almost as hard as approximating Densest $\kappa$-subgraph problem (that is, if there is a $\rho$ approximation algorithm for SSVE then there is a $(2\rho^2)$-approximation algorithm for the Densest $\kappa$-subgraph problem). This result suggests that although the SSVE problem looks similar to the Small Set Expansion (SSE) problem [RS10], it might be much harder than SSE. On the other hand, the SSVE problem is of independent interest – besides its application to the (st)-VC-kRC problem, Applebaum et al. [ABW10] used a "planted" version of SSVE as a hardness assumption to construct a public key encryption scheme.

**Other Related Work**  Another version of the EC-kRC problem that has received a significant amount of attention recently is the single-source setting. In this setting we are given a single source $s$ and a set $T$ of $r$ terminals. The source-sink pairs are then set to be $\{(s,t)\}_{t \in T}$. Bruhn et al. [BČH+08] have shown a factor $2(k-1)$-approximation for the uniform version of this problem, and Barman and Chawla [BC10] have shown a factor $(6, O(\sqrt{r} \ln r))$-bi-criteria approximation for the general version, a factor $(4, 4)$-bi-criteria approximation for the general version where $r$ is a constant, and a factor $(2, 4)$-approximation for the uniform version and arbitrary $r$.

The (st)-EC-kRC and (st)-VC-kRC problems capture two natural budgetted cut minimization problems. The first is the Minimum Unbalanced cut problem [HKPS05], in which we are given a graph $G$ with a source vertex $s$ and a budget $B$. The goal is to find a cut $(S, \bar{S})$ in $G$ with $s \in S$ and $|E(S, \bar{S})| \leq B$, while minimizing $|S|$. Hayrapetyan et al. [HKPS05] obtain a $(1 + 1/\lambda, \lambda)$-bi-criteria approximation algorithm for any $\lambda > 1$, by rounding a Lagrangean relaxation for the problem. Given an instance $G = (V, E)$ of the Minimum Unbalanced cut problem, we can transform it into an instance of (st)-EC-kRC, by setting the weights of all edges in $E$ to $\infty$, adding a sink $t$, that connects to every vertex in $V$ with a unit-weight edge, and setting $k = B$. The other problem is the Minimum $k$-size (st)-cut problem, where we are given a graph $G = (V, E)$ with a special source vertex $s$ and a parameter $k$, and the goal is to find a cut $(S, \bar{S})$ in $G$ with $s \in S$ and $|S| \leq k$, minimizing the size of the cut $|E(S, \bar{S})|$. Li and Zhang [LZ10] give an $O(\log n)$-approximation to this problem using Räcke's graph decomposition [Räc08]. This problem can be reduced to (st)-EC-kRC by assigning unit weights to the edges of $E$, and adding a sink $t$ with infinity-weight edges $(v, t)$ for each $v \in V$; the parameter $k$ remains unchanged.

### Our results and techniques

The following two theorems summarize our results for the EC-kRC problem.

**Theorem 1.1** *There is an efficient $O(k \log^{1.5} r)$-approximation algorithm, and a $\left(1 + \delta, O\left(\frac{1}{\delta} \log^{1.5} r\right)\right)$-bi-criteria approximation algorithm for any constant $\delta \in (0, 1)$, for the uniform EC-kRC problem.*

**Theorem 1.2** *There is a $\left(2, O(\log^{2.5} r \log \log r)\right)$-bi-criteria approximation algorithm with running time $n^{O(k)}$ and an $\left(O(\log r), O(\log^3 r)\right)$-bi-criteria approximation algorithm with running time $\text{poly}(n)$ for the EC-kRC problem.*

We now proceed to discuss our techniques. Our algorithms are based on a simple iterative approach: find a "sparse" cut that separates some demand pairs, remove all cut edges except for the $(k-1)$ most expensive ones from the graph, also remove all demand pairs that are no longer $k$-connected, and then recursively solve the obtained instance. The main challenge in this approach is to ensure that the cost of the removed edges is bounded by the cost of the optimal solution. In fact, in the first step of the algorithm, we use a modified notion of sparsity — we use the $k$-route sparsity of a cut, which is the cost of all but $(k-1)$ most expensive edges of the cut divided by the number of separated terminals (see below for formal definitions). This is necessary since the standard sparsest cut can be prohibitively expensive; its cost cannot be bounded in terms of the cost of the optimal solution. We prove however that the cost of the $k$-route sparsest cut can be bounded in terms of the cost of the optimal solution and thus obtain guarantees on the performance of our algorithms. This is the most technically challenging part of the analysis of our algorithms.

We extend our bi-criteria approximation for EC-kRC to the more general VC-kRC problem in the following theorem.

**Theorem 1.3** *There is a $\left(2, O(dk \log^{2.5} r \log \log r)\right)$-bi-criteria approximation algorithm for VC-kRC with running time $n^{O(k)}$, where $d$ is the maximum number of demand pairs in which any terminal participates.*

We also prove the following hardness of approximation result for VC-kRC, whose proof uses ideas similar to those used by Kortsarz et al. [KKL04] and Chakraborty et al. [CCK08] to prove hardness of vertex-connectivity network design:

**Theorem 1.4** *There are constants $0 < \epsilon < 1$, $k_0 > 1$, such that for any constant $\eta$, for any $k = O\left(2^{(\log n)^{1-\eta}}\right)$, where $k > k_0$, there is no $k^\epsilon$-approximation algorithm for VC-kRC, under the assumption that $\mathsf{P} \neq \mathsf{NP}$ for constant $k$, and under the assumption that $\mathsf{NP} \nsubseteq \mathsf{DTIME}(n^{\text{poly} \log n})$ for super-constant $k$.*

Finally, for the special case of $k = 2$, we obtain a slightly improved approximation algorithm:

**Theorem 1.5** *There is an efficient factor $O(\log^{1.5} r)$-approximation algorithm for both VC-kRC and EC-kRC, when $k = 2$.*

We now turn to the single (st)–pair version of the problems. We start with a simple approximation algorithm, summarized in the next theorem.

**Theorem 1.6** *There is an efficient factor $(k+1)$-approximation algorithm, and for every constant $c > 0$, there is an efficient $\left(1 + \frac{1}{c}, 1 + c\right)$-bi-criteria approximation algorithm for both (st)-VC-kRC and (st)-EC-kRC.*

We then proceed to show inapproximability results for the single (st)–pair version of the problem. Our first inapproximability result uses Feige's random $\kappa$-AND assumption [Fei02]. Given parameters $n$ and $\Delta$, a random $\kappa$-AND instance is defined to be a $\kappa$-AND formula on $n$ variables and $m = \Delta n$ clauses, where each clause chooses $\kappa$ literals uniformly at random from the set of $2n$ available literals. We say that a formula $\Phi$ is $\alpha$-satisfiable iff there is an assignment to the variables that satisfies an $\alpha$-fraction of the clauses. Notice that a random assignment satisfies a $1/2^\kappa$-fraction of the clauses in expectation, and we expect that this is a typical number of simultaneously satisfiable clauses for a random $\kappa$-AND formula. We next state the Random $\kappa$-AND conjecture of Feige [Fei02] and our inapproximability result for (st)-VC-kRC.

**Hypothesis 1.1** *(Random $\kappa$-AND assumption: Hypothesis 3 in [Fei02]). For some constant $c_0 > 0$, for every $\kappa$, there is a value of $\Delta_0$, such that for every $\Delta > \Delta_0$, there is no polynomial time algorithm that for random $\kappa$-AND formulas with $n$ variables and $m = \Delta n$ clauses, outputs 'typical' with probability $1/2$, but never outputs 'typical' on instances with $m/2^{c_0\sqrt{\kappa}}$ simultaneously satisfiable clauses.*

**Theorem 1.7** *For every constant $C > 0$, there exists a small constant $0 < \gamma < 1$ which depends only on $C$, such that assuming Hypothesis 1.1, there is no polynomial time algorithm which obtains a $(1 + \gamma, C)$-bi-criteria approximation for the (st)–VC-kRC problem.*

We also prove a slightly different inapproximability result based on the slightly weaker Random 3-SAT assumption of Feige. Given parameters $n$ and $\Delta$, a random 3-SAT formula on $n$ variables and $m = \Delta n$ clauses is constructed as follows: each clause chooses 3 literals uniformly at random among all available literals. Notice that a random assignment satisfies a 7/8-fraction of clauses in expectation. Below is Feige's 3-SAT assumption and our inapproximability result for (st)-VC-kRC.

**Hypothesis 1.2** *(Random 3-SAT assumption: Hypothesis 2 from [Fei02]). For every fixed $\epsilon > 0$, for $\Delta$ a sufficiently large constant independent of $n$, there is no polynomial time algorithm that on a random 3CNF formula with $n$ variables and $m = \Delta n$ clauses, outputs 'typical' with probability at least $1/2$, but never outputs 'typical' when the formula is $(1-\epsilon)$-satisfiable (i.e. there is an assignment satisfying simultaneously $(1-\epsilon)m$ clauses).*

**Theorem 1.8** *Assuming Hypothesis 1.2, for any constant $\epsilon > 0$, no polynomial-time algorithm achieves a $\left(\frac{25}{24} - \epsilon, 1.1 - \epsilon\right)$-bi-criteria approximation for (st)–VC-kRC.*

Finally, we show that an existence of a good approximation algorithm for (st)-VC-kRC would imply a good approximation for the $\lambda$-uniform Hypergraph Densest $\kappa$-subgraph problem. Recall that in the $\lambda$-uniform Hypergraph Densest $\kappa$-subgraph problem, we are given a graph $G(V, E)$ where $E$ is the set of $\lambda$-uniform hyperedges, and a parameter $\kappa$. The goal is to find a subset $S \subseteq V(G)$ of $\kappa$ vertices, maximizing the number of hyper-edges $e \subseteq S$. Notice that for $\lambda = 2$, this is the standard Densest $\kappa$-subgraph problem.

**Theorem 1.9** *For any constant $\lambda \geq 2$, and for any approximation factor $\rho$ (that may depend on $n$), if there is an efficient factor $\rho$ approximation algorithm for the (st)–VC-kRC problem, then there is an efficient factor $(2\rho^\lambda)$-approximation algorithm for the $\lambda$-uniform Hypergraph Densest $\kappa$-subgraph problem.*

We note that Theorem 1.9, combined with the recent result of [AAM[+]11] immediately implies super-constant inapproximability for (st)-VC-kRC, under Hypothesis 1.1. However, our proof of Theorem 1.7 is conceptually simpler, and also leads to a bi-criteria inapproximability.

**Organization** We present notation and definitions and prove some results that we use throughout the paper in Section 2. We study the uniform case of EC-kRC in Section 3, and the non-uniform case in Section 4. We describe our results for VC-kRC in Section 5. Then we present an algorithm for 2-route cuts in Section 6. We prove $n^\epsilon$ hardness of VC-kRC in Section 7. Finally, we study the single source-sink case in Section 8, where we present an approximation algorithm and prove several hardness results for the problem.

## 2 Preliminaries

In all our problems, the input is an undirected $n$-vertex graph $G = (V, E)$ with non-negative weights $w(e)$ on edges $e \in E$ and a parameter $k$. Additionally, we are given a set $D = \{(s_1, t_1), \ldots, (s_r, t_r)\}$ of source-sink pairs, that we also refer to as demand pairs. We let $T \subseteq V$ be the subset of vertices that participate in any demand pairs, and we refer to the vertices in $T$ as terminals. For every vertex $v \in V$, let $D_v$ be the number of demand pairs in which $v$ participates. Given a subset $S \subseteq V$ of vertices, let $D(S) = \sum_{v \in S} D_v$ be the total number of terminals in $S$, counting multiplicities. We also denote by $D(S, \bar{S})$ the number of demand pairs $(s_i, t_i)$ with $s_i \in S$, $t_i \in \bar{S}$, or $s_i \in \bar{S}$ and $t_i \in S$. Given any subset $E' \subseteq E$ of edges, we denote by $w(E') = \sum_{e \in E'} w(e)$ its weight. Throughout the paper, we denote by $E^*$ the optimal solution to the given EC-kRC or VC-kRC problem instance, and by OPT its value.

One of the main ideas in our algorithms is to relate the value of the appropriately defined sparsest cut in graph $G$ to the value of the optimal solution to the $k$-route cut problem. We now define the different variations of the sparsest cut problem that we use.

**The Sparsest Cut Problem.** Given any cut $(S, \bar{S})$ in graph $G$, its *uniform sparsity* is defined to be
$$\Phi(S) = \frac{w(E(S, \bar{S}))}{\min\{D(S), D(\bar{S})\}}.$$

The uniform sparsity $\Phi(G)$ of the graph $G$ is the minimum sparsity of any cut in $G$,

$$\Phi(G) = \min_{\substack{S \subset V:\\ D(S), D(\bar{S}) > 0}} \{\Phi(S)\}.$$

We use the $O(\sqrt{\log r})$-approximation algorithm for the uniform sparsest cut problem due to Arora, Rao and Vazirani [ARV04]. Let $\mathcal{A}_{\mathrm{ARV}}$ denote this algorithm, and let $\alpha_{\mathrm{ARV}}(r) = O(\sqrt{\log r})$ denote its approximation factor. Given an edge-weighted graph $G$ and a set $D$ of $r$ demand pairs, algorithm $\mathcal{A}_{\mathrm{ARV}}$ finds a subset $S \subseteq V$ of vertices with $\Phi(S) \leq \alpha_{\mathrm{ARV}}(r) \cdot \Phi(G)$.

Given any cut $(S, \bar{S})$ in graph $G$, its *non-uniform sparsity* is defined to be

$$\tilde{\Phi}(S) = \frac{w(E(S, \bar{S}))}{D(S, \bar{S})}.$$

The non-uniform sparsity $\tilde{\Phi}(G)$ of the graph $G$ is:

$$\tilde{\Phi}(G) = \min_{\substack{S \subset V:\\ D(S,\bar{S}) > 0}} \{\tilde{\Phi}(S)\}.$$

We also use the $O(\sqrt{\log r} \cdot \log \log r)$-approximation algorithm for the non-uniform sparsest cut problem of Arora, Lee and Naor [ALN05]. Let $\mathcal{A}_{\mathrm{ALN}}$ denote this algorithm, and let $\alpha_{\mathrm{ALN}}(r) = O(\sqrt{\log r} \cdot \log \log r)$ denote its approximation factor. Given an edge-weighted graph $G$ with a set $D$ of $r$ demand pairs, algorithm $\mathcal{A}_{\mathrm{ALN}}$ finds a subset $S \subseteq V$ of vertices with $\tilde{\Phi}(S) \leq \alpha_{\mathrm{ALN}}(r) \cdot \tilde{\Phi}(G)$.

We next generalize the notion of the sparsest cut to the multi-route setting. Given a subset $S \subseteq V$ of vertices, let $F$ denote the set of $(k-1)$ most expensive edges of $E(S, \bar{S})$, breaking ties arbitrarily, and we refer to $F$ as the set of *free edges* for cut $(S, \bar{S})$. We then define $w^{(k)}(S, \bar{S}) = \sum_{e \in E(S,\bar{S}) \setminus F} w_e$.

The *uniform k-route sparsity* of set $S$ is defined to be:

$$\Phi^{(k)}(S) = \frac{w^{(k)}(S, \bar{S})}{\min\{D(S), D(\bar{S})\}},$$

and the uniform $k$-route sparsity of the graph $G$ is:

$$\Phi^{(k)}(G) = \min_{\substack{S \subset V:\\ D(S), D(\bar{S}) > 0}} \{\Phi^{(k)}(S)\}.$$

Similarly, the *non-uniform k-route sparsity* of $S$ is:

$$\tilde{\Phi}^{(k)}(S) = \frac{w^{(k)}(S, \bar{S})}{D(S, \bar{S})},$$

and the non-uniform $k$-route sparsity of the graph $G$ is:

$$\tilde{\Phi}^{(k)}(G) = \min_{\substack{S \subset V:\\ D(S,\bar{S}) > 0}} \{\tilde{\Phi}^{(k)}(S)\}.$$

Note that $\Phi^{(1)}(G) = \Phi(G)$ and $\tilde{\Phi}^{(1)}(G) = \tilde{\Phi}(G)$ are the standard uniform and non-uniform sparsest cut values, respectively. We now show that there is an efficient algorithm to find an approximate $k$-route sparsest cut when $k$ is a constant.

**Theorem 2.1** *There is an algorithm that, given a graph $G = (V, E)$ with $r$ source-sink pairs and an integer $k$, computes in time $n^{O(k)}$ a cut $S \subseteq V$, with $\Phi^{(k)}(S) \leq \alpha_{\mathrm{ARV}}(r) \cdot \Phi^{(k)}(G)$. Similarly, there is an algorithm that computes in time $n^{O(k)}$ a cut $S$, with $\tilde{\Phi}^{(k)}(S) \leq \alpha_{\mathrm{ALN}}(r) \cdot \tilde{\Phi}^{(k)}(G)$.*

**Proof:** We start with the uniform $k$-route sparsest cut. We go over all subsets $F \subseteq E$ of $k - 1$ edges. For each such subset $F$, we compute the $\alpha_{\mathrm{ARV}}(r)$-approximate sparsest cut in the graph $G \setminus F$ using the algorithm $\mathcal{A}_{\mathrm{ARV}}$, and output the best cut over all such subsets $F$. The algorithm for the non-uniform sparsest $k$-route cut is similar, except that we use the algorithm $\mathcal{A}_{\mathrm{ALN}}$ for the non-uniform sparsest cut. $\square$

The above theorem works well for constant values of $k$. However, when $k$ is super-constant, the running time of the algorithm is no longer polynomial. For such cases, we use a bi-criteria approximation algorithm for the $k$-route sparsest cut problem, summarized in the next theorem.

**Theorem 2.2** *There is an efficient algorithm that, given an edge-weighted graph $G = (V, E)$, an integer $k > 1$, and a set $D = \{(s_i, r_i)\}_{i=1}^{r}$ of $r$ demand pairs, finds a cut $S \subseteq V$ with $\tilde{\Phi}^{(k')}(S) = O(\log r) \cdot \tilde{\Phi}^{(k)}(G)$, where $k' = Ck \log r$ for some absolute constant $C$.*

**Proof:** We use as a sub-routine the approximation algorithm of Englert et al. [EGK$^+$10] for the $\ell$-Multicut problem. In the $\ell$-Multicut problem, we are given a graph $G = (V, E)$ with weights on edges, a set $D$ of $r$ demand pairs, and an integer $\ell$. The goal is to find a minimum-weight subset $E' \subseteq E$ of edges, such that at least $\ell$ of the demand pairs are disconnected in the graph $G \setminus E'$. Engert et al. [EGK$^+$10] give an efficient $O(\log r)$-approximation algorithm for this problem. We denote their algorithm by $\mathcal{A}_{\mathrm{EGK+}}$, and the approximation factor it achieves by $\alpha_{\mathrm{EGK+}} = O(\log r)$.

Let $(S^*, \bar{S}^*)$ be the optimal non-uniform sparsest $k$-route cut in $G$, and let $F^* \subseteq E_G(S^*, \bar{S}^*)$ be the subset of the $(k-1)$ most expensive edges in this cut. Then $w(E(S^*, \bar{S}^*) \setminus F^*) = \tilde{\Phi}^{(k)}(G) \cdot D(S^*, \bar{S}^*)$. Let $W^* = w(E(S^*, \bar{S}^*) \setminus F^*)$ and let $r^* = D(S^*, \bar{S}^*)$.

Assume first that our algorithm is given the values of $W^*$ and $r^*$. We define new edge weights as follows: for each edge $e \in E$, $\tilde{w}_e = \min\{w_e, W^*/(k-1)\}$. We use the algorithm $\mathcal{A}_{\mathrm{EGK+}}$ on the resulting instance of the $\ell$-Multicut problem, with $\ell = r^*$. Let $E'$ be the output of the algorithm, and let $\mathcal{C}$ be the collection of connected components in $G \setminus E'$. We can then find a partition $(S, \bar{S})$ of the vertices of $G$, such that $E(S, \bar{S}) \subseteq E'$, and $D(S, \bar{S}) \geq r^*/2$, as follows. We start with an arbitrary partition $(S, \bar{S})$ of the vertices of $G$, where each cluster $C \in \mathcal{C}$ is contained in either $S$ or $\bar{S}$. We then perform a number of iterations. In each iteration, if there is a cluster $C \in \mathcal{C}$, such that moving all its vertices to the opposite side of the current cut increases $D(S, \bar{S})$, we move the vertices of $C$ to the opposite side of the cut. It is easy to verify that in the final partition $(S, \bar{S})$, $D(S, \bar{S}) \geq r^*/2$.

Let $(S, \bar{S})$ be the resulting partition, and let $F$ be the set of $2\alpha_{\mathrm{EGK+}}(r)(k-1)$ most expensive edges of $E(S, \bar{S})$, with respect to the original weights $w_e$, breaking ties arbitrarily.

The output of our algorithm is the cut $(S, \bar{S})$. In order to complete the proof, it is enough to show that $w(E(S, \bar{S}) \setminus F) \leq O(\log r) \cdot \tilde{\Phi}^{(k)}(G) \cdot D(S, \bar{S})$.

Note that the value of the optimal solution to the $\ell$-Multicut problem instance is at most

$$\tilde{w}(E(S^*, \bar{S}^*)) \leq \tilde{w}(E(S^*, \bar{S}^*) \setminus F^*) + |F^*| \cdot \frac{W^*}{k-1}$$

$$\leq w(E(S^*, \bar{S}^*) \setminus F^*) + W^* = 2W^*.$$

Therefore, $\tilde{w}(E(S, \bar{S})) \leq 2\alpha_{\text{EGK+}}(r)W^*$. In particular, $E(S, \bar{S})$ may contain at most $2\alpha_{\text{EGK+}}(r)(k-1)$ edges $e$ with $\tilde{w}_e = W^*/(k-1)$, and so all such edges lie in $F$. For edges $e \notin F$, $\tilde{w}_e < W^*/(k-1)$ must hold, and therefore, $\tilde{w}_e = w_e$. We conclude that

$$w(E(S, \bar{S}) \setminus F) = \tilde{w}(E(S, \bar{S}) \setminus F) \leq \tilde{w}(E(S, \bar{S}))$$
$$\leq 2\alpha_{\text{EGK+}}(r)W^* = 2\alpha_{\text{EGK+}}(r)\tilde{\Phi}^{(k)}(G) \cdot r^*$$
$$\leq O(\log r)\tilde{\Phi}^{(k)}(G)D(S, \bar{S})$$

as required.

Of course, our algorithm does not know the values of $W^*$ and $r^*$. Instead, we perform the procedure described above for all possible values of $r' \in \{1, \ldots, r\}$ and (say) all values of $W'$ in $\{\tau w_e : e \in E, 1 \leq \tau \leq |E|\}$, and then output the best of the cuts found. One of the values of $r'$ will be equal to $r^*$, and one the values of $W'$ will be within a factor of 2 of $W^*$: if $e$ is the most expensive edge in $E(S^*, \bar{S}^*) \setminus F^*$, and $\tau = \lceil W^*/w_e \rceil$, then $W^* \leq \tau w_e \leq (2\lfloor W^*/w_e \rfloor)w_e \leq 2W^*$. For these values of $r'$ and $W'$, the algorithm will find a cut that satisfies the conditions of the lemma. □

**Laminar Families of Minimum Cuts** Our main tool in establishing the connection between the values of the $k$-route sparsest cut and the cost of the optimal solution to the $k$-route cut problem is the following theorem, which shows that there is a laminar family of minimum cuts disconnecting the source-sink pairs in the graph $G$.

**Lemma 2.3** *There is an efficient algorithm, that, given any edge-weighted graph $G = (V, E)$ with a set $D = \{(s_i, t_i)\}_{i=1}^r$ of $r$ source-sink pairs, finds a laminar family $\mathcal{S} = \{S_1, \cdots, S_r\}$ of vertex subsets, such that for all $1 \leq i \leq r$:*

- *$(S_i, V \setminus S_i)$ is a minimum cut separating $s_i$ from $t_i$ in $G$, and*
- *$D(S_i) \leq r$ (so $S_i$ contains at most half the terminals, counting multiplicities).*

**Proof:** We use a Gomory–Hu tree $T_{GH}$ for the graph $G$. Recall that it is a weighted tree, whose vertex set is $V$. Let $c_e$ denote the costs of the edges $e \in E(T_{GH})$. Tree $T_{GH}$ has the following key property: for every pair $(u, v) \in V$ of vertices, the value of the minimum cut separating $u$ from $v$ in graph $G$ equals the value of the minimum cut separating $u$ from $v$ in $T_{GH}$. Note that the latter cut contains only one edge – the minimum-cost edge on the unique path connecting $u$ to $v$ in the tree.

We start with a Gomory-Hu tree $T_{GH}$ for the graph $G$. For each $1 \leq i \leq r$, let $(L_i, R_i)$ be a minimum cut separating $s_i$ from $t_i$ in $T_{GH}$. If $D(L_i) < D(R_i)$, then we set $S_i = L_i$. If $D(R_i) < D(L_i)$, we set $S_i = R_i$. Otherwise, if $D(R_i) = D(L_i)$, we let $S_i$ to be the side containing the vertex $s_1$. We use this tie-breaking rule that enforces consistency across different source-sink pairs later.

This finishes the definition of the family $\mathcal{S} = \{S_1, \ldots S_r\}$ of vertex subsets. It is immediate to see that for each $1 \leq i \leq r$, $(S_i, V \setminus S_i)$ is a minimum cut separating $s_i$ from $t_i$ in $G$, and that $D(S_i) \leq r$. It now only remains to show that $S_1, \ldots, S_r$ form a laminar family.

Assume for contradiction that for some $i \neq j$, $S_i \cap S_j \neq \emptyset$, but $S_i \setminus S_j \neq \emptyset$, and $S_j \setminus S_i \neq \emptyset$. Let $e_i$ be the unique edge of $T_{GH}$ lying in the cut $(S_i, V \setminus S_i)$ in tree $T_{GH}$, and let $e_j$ be the unique edge of $T_{GH}$ lying in the cut $(S_j, V \setminus S_j)$. Observe that $T_{GH} \setminus \{e_i, e_j\}$ consists of three non-empty connected components. Let $C_1$ denote the component that is incident on both $e_i$ and $e_j$, $C_2$ the component incident on $e_j$ only, and $C_3$ the remaining component. We claim that $S_i = C_1 \cup C_2$. Otherwise, since edge $e_i$ separates $S_i$ from $V \setminus S_i$ in $T_{GH}$, $S_i = C_3$ must hold. But then either $S_j = C_2$ and so $S_i \cap S_j = \emptyset$, or $S_j = C_1 \cup C_3$ and then $S_i \subseteq S_j$, a contradiction. Therefore, $S_i = C_1 \cup C_2$ and $V \setminus S_i = C_3$. Similarly, $S_j = C_1 \cup C_3$ and $V \setminus S_j = C_2$.

From the definition of $S_i$, either $D(S_i) < D(V \setminus S_i)$, or $D(S_i) = D(V \setminus S_i)$ and $s_1 \in S_i$. Assume first that $D(S_i) < D(V \setminus S_i)$. Then $V \setminus S_j = C_2 \subseteq S_i$, and so $D(V \setminus S_j) < D(S_j)$, contradicting the definition of $S_j$. We reach a similar contradiction if $D(S_j) < D(V \setminus S_j)$. Therefore, $D(S_i) = D(V \setminus S_i)$ and $D(S_j) = D(V \setminus S_j)$ must hold. In other words, $D(V \setminus S_i) = D(C_3) = r$, and $D(V \setminus S_j) = D(C_2) = r$. Since $C_2$ and $C_3$ are disjoint, this means that $D(C_1) = 0$. But from the definitions of $S_i$ and $S_j$, $s_1 \in S_i \cap S_j$ must hold, a contradiction. □

**First Algorithmic Framework** Most our algorithms belong to one of two simple algorithmic frameworks. The first framework uses a divide-and-conquer paradigm: We start with the graph $G = (V, E)$ and a set $D$ of $r \geq 1$ demand pairs, and then find a cut $(S, \bar{S})$ in $G$, with $D(S), D(\bar{S}) \geq 1$. We then select a subset $E_0 \subseteq E(S, \bar{S})$ of edges to delete, and apply the algorithm recursively to the sub-instances induced by $G[S]$ and $G[\bar{S}]$. Here, the sub-instance induced by $G[S]$ consists of the graph $G[S]$ and the collection of the original demand pairs $(s_i, t_i)$, with both $s_i, t_i \in S$. The sub-instance induced by $G[\bar{S}]$ is defined similarly. Let $E_1$ and $E_2$ be the solutions returned by the two recursive calls, respectively. The final solution is $E' = E_0 \cup E_1 \cup E_2$. The specific cut $(S, \bar{S})$, and the subsets $E_0 \subseteq E(S, \bar{S})$ of edges computed will differ from algorithm to algorithm, and we will need to select them in a way that ensures the feasibility of the final solution. However, the analysis of the solution cost is similar in all these algorithms, and is summarized in the following theorem.

**Theorem 2.4** *Let $\mathcal{A}$ be any algorithm in the above framework, and assume that the algorithm guarantees that $w(E_0) \leq \frac{\alpha \cdot \mathsf{OPT}}{r} \cdot \min\{D(S), D(\bar{S})\}$, for some factor $\alpha$. Then $w(E') \leq 4\alpha \ln(1 + r) \cdot \mathsf{OPT}$.*

**Proof:** The proof is by induction on $r$. If $r = 1$ then $E' = E_0$, and the statement trivially holds. Assume now that the statement holds for instances with fewer than $r$ demand pairs, for some $r > 1$. Consider the cut $(S, \bar{S})$ computed by the algorithm $\mathcal{A}$ on the current instance. Let $a$ be the number of demand pairs $(s_i, t_i)$ with $s_i, t_i \in S$, let $b$ be the number of demand pairs $(s_i, t_i)$ with $s_i, t_i \in \bar{S}$, and assume w.l.o.g. that $a \geq b$. Then $D(S) \geq 2a$, $D(\bar{S}) \geq 2b$, and so $D(S) \leq 2r - D(\bar{S}) \leq 2(r - b)$ and $D(\bar{S}) \leq 2(r - a)$. Therefore,

$$w(E_0) \leq \frac{\alpha \cdot \mathsf{OPT}}{r} \cdot \min\{D(S), D(\bar{S})\} \leq \frac{2\alpha \cdot \mathsf{OPT}}{r}(r - a).$$

The optimal solutions to the EC-kRC instances on graphs $G[S]$ and $G[\bar{S}]$ have costs at most $w(E^* \cap E(S))$ and $w(E^* \cap E(\bar{S}))$, respectively. By the induction hypothesis, the total cost of solutions $E_1$ and $E_2$ on graphs $G[S]$ and $G[\bar{S}]$ is at most

$$4\alpha w(E^* \cap E(S)) \ln(1+a) + 4\alpha w(E^* \cap E(\bar{S})) \ln(1+b)$$
$$\leq 4\alpha \Big(w(E^* \cap E(S)) + w(E^* \cap E(\bar{S}))\Big) \ln(1+a)$$
$$\leq 4\alpha \cdot \mathsf{OPT} \cdot \ln(1+a).$$

The total solution cost is then bounded by:

$$w(E') \leq 4\alpha \cdot \mathsf{OPT} \cdot \ln(1+a) + \frac{2\alpha \cdot \mathsf{OPT}}{r}(r-a)$$
$$\leq 4\alpha \cdot \mathsf{OPT} \left(\ln(1+a) + \frac{r-a}{2r}\right).$$

The theorem follows from the following inequality:

$$\ln(1+a) + \frac{r-a}{2r} = \ln(1+r) + \ln\left(\frac{1+a}{1+r}\right) + \frac{r-a}{2r}$$
$$\leq \ln(1+r) - \frac{r-a}{1+r} + \frac{r-a}{2r}$$
$$\leq \ln(1+r),$$

where we have used the fact that $\ln\left(\frac{1+a}{1+r}\right) = \ln\left(1 - \frac{r-a}{1+r}\right) \leq -\frac{r-a}{1+r}$, since $\ln(1+x) \leq x$ for all $x > -1$. $\square$

**Second Algorithmic Framework** The algorithmic framework presented above has some limitations. Specifically, we can only use it in scenarios where there is a cheap collection $E'$ of edges (with cost roughly comparable to $\mathsf{OPT}$), whose removal decomposes our instance $G$ into two disjoint sub-instances, $G[S], G[\bar{S}]$, which can then be solved separately. This is the case for the uniform EC-kRC, and the non-uniform EC-kRC and VC-kRC when $k = 2$. For higher values of $k$ in the non-uniform setting, such a decomposition may not exist. Instead, we use the following framework. Given a graph $G$ and a set $D$ of $r \geq 1$ demand pairs, we find a collection $E_0$ of edges to delete, together with a subset $D_0$ of demand pairs to remove, where $|D_0| \geq 1$. We then solve the problem recursively on the graph $G' = G \setminus E_0$, and the set $D \setminus D_0$ of the remaining demand pairs. Let $E_1$ be the subset of edges returned by the recursive call. Then the solution computed by the algorithm is $E' = E_0 \cup E_1$. The specific subset $E_0$ of edges to remove and the subset $D_0$ of demands will again be computed differently by each algorithm, in a way ensures that the final solution is feasible. The analysis of the solution cost of such algorithms is summarized in the next theorem.

**Theorem 2.5** *Let $\mathcal{A}$ be any algorithm in the above framework, and assume that we are guaranteed that $w(E_0) \leq \alpha \cdot \mathsf{OPT} \cdot \frac{|D_0|}{r}$, for some factor $\alpha$. Then $w(E') \leq 2\alpha \ln(1+r) \cdot \mathsf{OPT}$.*

**Proof:** The proof is by induction on $r$. If $r = 1$ then $E' = E_0$, and the statement trivially holds. Assume now that the statement holds for instances with fewer than $r$ demand pairs, for some $r > 1$.

We prove the theorem for instances with $r$ demand pairs. Let $a = |D_0|$. Then by the induction hypothesis, $w(E_1) \leq 2\alpha \cdot \mathsf{OPT} \cdot \ln(1 + r - a)$. Therefore,

$$\begin{aligned}
w(E') &\leq 2\alpha \cdot \mathsf{OPT} \cdot \ln(1 + r - a) + \alpha \cdot \mathsf{OPT} \cdot \frac{a}{r} \\
&= 2\alpha \cdot \mathsf{OPT} \left( \ln(r+1) + \ln\left(\frac{1+r-a}{r+1}\right) + \frac{a}{2r} \right) \\
&\leq 2\alpha \cdot \mathsf{OPT} \left( \ln(r+1) - \frac{a}{r+1} + \frac{a}{2r} \right) \\
&\leq 2\alpha \ln(r+1) \mathsf{OPT}.
\end{aligned}$$

$\square$

## 3 Uniform EC-kRC

This section is dedicated to proving Theorem 1.1. We first show an $O(k \log^{1.5} r)$-approximation algorithm, and provide a bi-criteria algorithm later. Recall that we are given an unweighted graph $G = (V, E)$, a set $\{(s_i, t_i)\}_{i=1}^{r}$ of demand pairs, and an integer $k$. Our goal is to find a collection $E'$ of $O(k \log^{3/2} r) \cdot \mathsf{OPT}$ edges, such that for each demand pair $(s_i, t_i)$, there are at most $(k-1)$ edge-disjoint paths in graph $G \setminus E'$ connecting them.

We assume w.l.o.g. that each source-sink pair $(s_i, t_i)$ is $k$-edge connected in the current graph $G$. Our algorithm views the graph $G$ as an instance of the uniform sparsest cut problem. We use the algorithm $\mathcal{A}_{\mathrm{ARV}}$ to find a partition $(S, \bar{S})$ of $V$ with $\Phi(S) \leq \alpha_{\mathrm{ARV}}(r) \cdot \Phi(G)$, add the edges in $E(S, \bar{S})$ to the solution $E'$, and delete the demand pairs $(s_i, t_i)$ that are no longer $k$-edge connected from the list of source-sink pairs. Notice that each remaining source-sink pair must be contained either in $S$ or in $\bar{S}$. We then recursively solve the EC-kRC problem on the sub-instances induced by $G[S]$ and $G[\bar{S}]$. The algorithm is summarized in Figure 1.

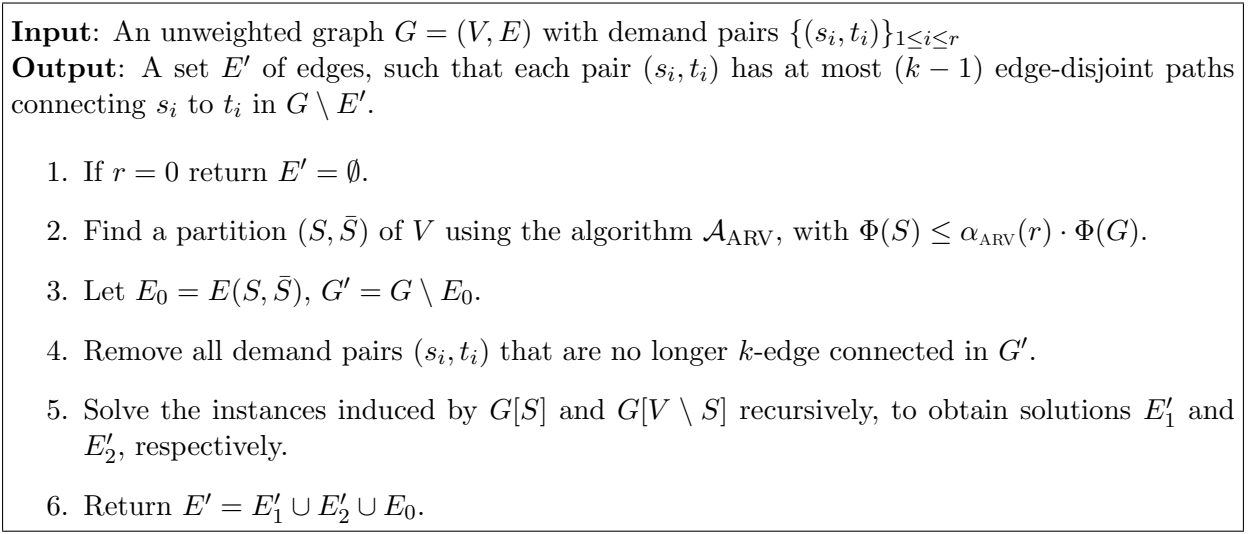

**Input**: An unweighted graph $G = (V, E)$ with demand pairs $\{(s_i, t_i)\}_{1 \leq i \leq r}$
**Output**: A set $E'$ of edges, such that each pair $(s_i, t_i)$ has at most $(k-1)$ edge-disjoint paths connecting $s_i$ to $t_i$ in $G \setminus E'$.

1. If $r = 0$ return $E' = \emptyset$.

2. Find a partition $(S, \bar{S})$ of $V$ using the algorithm $\mathcal{A}_{\mathrm{ARV}}$, with $\Phi(S) \leq \alpha_{\mathrm{ARV}}(r) \cdot \Phi(G)$.

3. Let $E_0 = E(S, \bar{S})$, $G' = G \setminus E_0$.

4. Remove all demand pairs $(s_i, t_i)$ that are no longer $k$-edge connected in $G'$.

5. Solve the instances induced by $G[S]$ and $G[V \setminus S]$ recursively, to obtain solutions $E_1'$ and $E_2'$, respectively.

6. Return $E' = E_1' \cup E_2' \cup E_0$.

Figure 1: Approximation algorithm for uniform EC-kRC.

The heart of the analysis of the algorithm is the following theorem, that relates the value of the uniform sparsest cut in graph $G$ to the value $\mathsf{OPT}$ of the optimal solution for EC-kRC.

**Theorem 3.1** *Suppose that we are given an unweighted graph $G = (V, E)$ with $r$ source-sink pairs $\{(s_i, t_i)\}_{i=1}^r$, such that for each pair $(s_i, t_i)$, there are at least $k$ edge-disjoint paths connecting $s_i$ to $t_i$ in $G$, and let $\mathsf{OPT}$ be the cost of the optimal solution of EC-kRC on $G$. Then $\Phi(G) \leq \frac{2k}{r} \cdot \mathsf{OPT}$.*

**Proof:** Consider the graph $H = G \setminus E^*$. We use Lemma 2.3 with edge weights $w_e = 1$ on graph $H$ to obtain the laminar family $\mathcal{S} = \{S_i\}_{i=1}^r$ of vertex subsets. Consider all maximal sets in the laminar family, that is, sets $S_i$ that are not contained in other sets. Assume w.l.o.g. that these sets are $S_1, \ldots, S_q$, for some $q \leq r$. Then $\sum_{i=1}^q D(S_i) \geq r$ must hold. Note that for each $i$, $|E_H(S_i, V \setminus S_i)| \leq k - 1$ since $s_i$ and $t_i$ are not $k$-edge connected in $H$, and $(S_i, V \setminus S_i)$ is a minimum cut separating $s_i$ from $t_i$ in $H$. On the other hand, $|E_G(S_i, V \setminus S_i)| \geq k$ since $s_i$ and $t_i$ are $k$-edge connected in $G$. Therefore,

$$\begin{aligned} |E_G(S_i, V \setminus S_i)| &= |E_H(S_i, V \setminus S_i)| \\ &\quad + |E_G(S_i, V \setminus S_i) \cap E^*| \\ &\leq (k-1) + |E_G(S_i, V \setminus S_i) \cap E^*| \\ &\leq k |E_G(S_i, V \setminus S_i) \cap E^*|. \end{aligned} \quad (1)$$

Note that every edge $e \in E^*$ belongs to at most two cuts $E_G(S_i, V \setminus S_i)$ and $E_G(S_j, V \setminus S_j)$. Therefore,

$$\begin{aligned} \sum_{i=1}^q |E_G(S_i, V \setminus S_i)| &\leq \sum_{i=1}^q k |E_G(S_i, V \setminus S_i) \cap E^*| \\ &\leq 2k \cdot \mathsf{OPT} \end{aligned}$$

On the other hand,

$$\begin{aligned} \sum_{i=1}^q |E_G(S_i, V \setminus S_i)| &= \sum_{i=1}^q \Phi(S_i) \cdot D(S_i) \\ &\geq \sum_{i=1}^q \Phi(G) \cdot D(S_i) \\ &\geq r \cdot \Phi(G). \end{aligned}$$

We conclude that $\Phi(G) \leq 2k \cdot \mathsf{OPT}/r$. □

We now analyze the algorithm. Since the algorithm removes a demand pair $(s_i, t_i)$ only when $s_i$ and $t_i$ are no longer $k$-edge connected, and terminates when all demand pairs are removed, the algorithm is guaranteed to find a feasible solution to the problem. In order to bound the solution cost, note that

$$\begin{aligned} |E_0| &= \Phi(S) \cdot \min\{D(S), D(\bar{S})\} \\ &\leq \alpha_{\mathrm{ARV}}(r) \cdot \Phi(G) \cdot \min\{D(S), D(\bar{S})\} \\ &\leq \frac{2k\alpha_{\mathrm{ARV}}(r)}{r} \cdot \mathsf{OPT} \cdot \min\{D(S), D(\bar{S})\}. \end{aligned}$$

We can now use Theorem 2.4 with $\alpha = 2k\alpha_{\mathrm{ARV}}(r)$ to conclude that $|E'| = O(k\alpha_{\mathrm{ARV}}(r) \log r)\mathsf{OPT} = O(k \log^{3/2} r)\mathsf{OPT}$.

**Bi-criteria approximation algorithm** We now slightly modify the algorithm from Figure 1, to obtain a $(1+\delta, O(\frac{1}{\delta}\log^{1.5} r))$-bi-criteria approximation algorithm for any constant $0 < \delta < 1$. The algorithm works exactly as before, except that it removes a demand pair $(s_i, t_i)$ in step 4 iff $s_i$ and $t_i$ are no longer $(1+\delta)k$ edge-connected. We also assume w.l.o.g. that in the original instance $G$, every demand pair $(s_i, t_i)$ has at least $(1+\delta)k$ edge-disjoint paths connecting $s_i$ to $t_i$. As before, it is straightforward to verify that if $E'$ is the final solution produced by the algorithm, then each demand pair $(s_i, t_i)$ pair has fewer than $(1+\delta)k$ edge-disjoint paths connecting them in $G \setminus E'$. In order to bound the solution cost, we prove the following analogue of Theorem 3.1.

**Theorem 3.2** *Suppose that we are given an unweighted graph $G$ with $r$ demand pairs $\{(s_i, t_i)\}_{i=1}^{r}$, where for each pair $(s_i, t_i)$, there are at least $(1+\delta)k$ edge-disjoint paths connecting $s_i$ to $t_i$ in $G$. Then $\Phi(G) \leq \frac{2\mathsf{OPT}}{r} \cdot (1 + 1/\delta)$.*

**Proof:** As before, we compute the laminar family of minimum cuts in graph $H = G \setminus E^*$, using Lemma 2.3, and we consider the collection of all maximal cuts in this family. Assume w.l.o.g. that it is $\{S_1, \ldots, S_q\}$, for $q \leq r$, and recall that $\sum_{i=1}^{q} D(S_i) \geq r$. As before, for each $1 \leq i \leq q$, $|E_G(S_i, V \setminus S_i)| \leq (k-1) + |E_G(S_i, V \setminus S_i) \cap E^*|$. Since $|E_G(S_i, V \setminus S_i)| \geq (1+\delta)k$, we get that $|E_G(S_i, V \setminus S_i) \cap E^*| \geq \delta k$, and so $(k-1) \leq |E_G(S_i, V \setminus S_i) \cap E^*|/\delta$. We get that:

$$|E_G(S_i, V \setminus S_i)| \leq (k-1) + |E_G(S_i, V \setminus S_i) \cap E^*|$$
$$\leq (1 + 1/\delta)|E_G(S_i, V \setminus S_i) \cap E^*|.$$

On the other hand, $|E_G(S_i, V \setminus S_i)| \geq \Phi(G) \cdot D(S_i)$. Summing up over all $1 \leq i \leq q$, we get that:

$$2\mathsf{OPT} \geq \sum_{i=1}^{q} |E_G(S_i, V \setminus S_i) \cap E^*|$$
$$\geq \frac{\delta}{\delta+1} \sum_{i=1}^{q} |E_G(S_i, V \setminus S_i)|$$
$$\geq \frac{\delta}{\delta+1} \Phi(G) \sum_{i=1}^{q} D(S_i) \geq \frac{\delta}{\delta+1} \Phi(G) \cdot r.$$

We conclude that $\Phi(G) \leq \frac{2\mathsf{OPT}}{r}(1 + 1/\delta)$. □

In order to bound the final solution cost, observe that

$$|E_0| = \Phi(S) \cdot \min\{D(S), D(\bar{S})\}$$
$$\leq \alpha_{\mathrm{ARV}}(r) \cdot \Phi(G) \cdot \min\{D(S), D(\bar{S})\}$$
$$\leq \frac{2\mathsf{OPT}\alpha_{\mathrm{ARV}}(r)}{r} \cdot (1 + 1/\delta) \cdot \min\{D(S), D(\bar{S})\}.$$

We now use Theorem 2.4 with $\alpha = 2\alpha_{\mathrm{ARV}}(r)(1+1/\delta)$ to conclude that $|E'| = O(\alpha_{\mathrm{ARV}}(r) \log r/\delta)\mathsf{OPT} = O(\log^{1.5} r/\delta)\mathsf{OPT}$, when $0 < \delta < 1$.

This concludes the proof of Theorem 1.1.

## 4 Non-uniform EC-kRC

In this section we prove Theorem 1.2. We start with a $(2,\tilde{O}(\log^{2.5} r))$-bi-criteria algorithm with running time $n^{O(k)}$, and we show an algorithm whose running time is polynomial in $n$ and $k$ later.

Abusing the notation, for each cut $(S,\bar{S})$ in graph $G$, we denote by $D(S,\bar{S})$ both the set of the demand pairs $(s_i,t_i)$ with $|\{s_i,t_i\}\cap S|=1$, and the number of such pairs.

### 4.1 A $(2,\tilde{O}(\log^{5/2} r))$ bi-criteria approximation in time $n^{O(k)}$

We cannot employ the first algorithmic framework for EC-kRC on weighted graphs. A natural approach in using it would be to find an appropriately defined sparse cut $(S,\bar{S})$, remove all but $k-1$ most expensive edges of this cut, and then recursively solve the problem on instances $G[S]$ and $G[\bar{S}]$. Let $E_0$ be the subset of edges removed, and let $G' = G \setminus E_0$ be the remaining graph. This approach does not work because it is possible that a demand pair $(s_i,t_i)$ with both $s_i,t_i \in S$ is connected by a path that visits $G[\bar{S}]$ in graph $G'$. So if we solve the problem recursively on $G[S]$ and $G[\bar{S}]$, then the combined solution is not necessarily a feasible solution to the problem. On the other hand, if, instead, we remove all or almost all edges of $E(S,\bar{S})$, then the resulting solution cost may be too high. Therefore, we employ the second algorithmic framework instead.

We assume w.l.o.g. that in the input graph $G$, each demand pair $(s_i,t_i)$ has at least $(2k-1)$ edge-disjoint paths connecting them. Our algorithm, summarized in Figure 2, starts by finding an approximate non-uniform $(2k-1)$-route sparse cut $(S,\bar{S})$ in $G$, using Theorem 2.1. That is, $\tilde{\Phi}^{(2k-1)}(S) \leq \alpha_{\text{ALN}}(r)\tilde{\Phi}^{(2k-1)}(G)$. Let $F$ be the set of the $(2k-2)$ most expensive edges of $E(S,\bar{S})$, let $E_0 = E(S,\bar{S}) \setminus F$, and let $G' = G \setminus E_0$. We remove all demand pairs that are no longer $(2k-1)$ connected in $G'$ (we denote the set of these demand pairs by $D_0$), and then recursively solve the resulting instance.

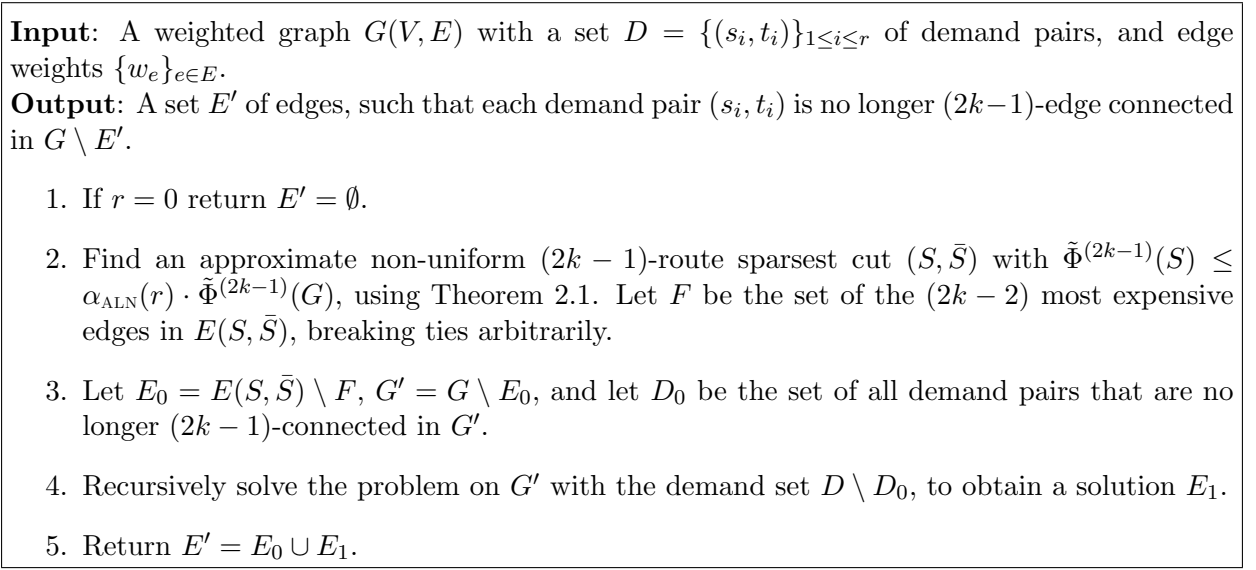

**Input**: A weighted graph $G(V,E)$ with a set $D = \{(s_i,t_i)\}_{1\leq i\leq r}$ of demand pairs, and edge weights $\{w_e\}_{e\in E}$.
**Output**: A set $E'$ of edges, such that each demand pair $(s_i,t_i)$ is no longer $(2k-1)$-edge connected in $G \setminus E'$.

1. If $r = 0$ return $E' = \emptyset$.

2. Find an approximate non-uniform $(2k-1)$-route sparsest cut $(S,\bar{S})$ with $\tilde{\Phi}^{(2k-1)}(S) \leq \alpha_{\text{ALN}}(r) \cdot \tilde{\Phi}^{(2k-1)}(G)$, using Theorem 2.1. Let $F$ be the set of the $(2k-2)$ most expensive edges in $E(S,\bar{S})$, breaking ties arbitrarily.

3. Let $E_0 = E(S,\bar{S}) \setminus F$, $G' = G \setminus E_0$, and let $D_0$ be the set of all demand pairs that are no longer $(2k-1)$-connected in $G'$.

4. Recursively solve the problem on $G'$ with the demand set $D \setminus D_0$, to obtain a solution $E_1$.

5. Return $E' = E_0 \cup E_1$.

Figure 2: A bi-criteria approximation algorithm for non-uniform EC-kRC in time $n^{O(k)}$.

It is immediate to verify that the algorithm returns a feasible solution. The running time of the algorithm is dominated by computing an approximate $(2k-1)$-route sparsest cut, and is therefore bounded by $n^{O(k)}$. In order to bound the solution cost, we use the following lemma that relates the value of $\tilde{\Phi}^{(2k-1)}(G)$ to $\mathsf{OPT}$.

**Theorem 4.1** *Suppose that we are given a graph $G(V,E)$ with edge weights $w_e$, and a set $D = \{(s_i,t_i)\}_{i\in[r]}$ of $r$ demand pairs, where every pair $(s_i,t_i)$ has at least $(2k-1)$ edge-disjoint paths connecting $s_i$ to $t_i$ in $G$. Then $\mathsf{OPT} = \Omega\left(\frac{r}{\log r}\right)\tilde{\Phi}^{(2k-1)}(G)$.*

**Proof:** Consider the graph $H = G \setminus E^*$. Let $\mathcal{S} = \{S_1,\ldots,S_r\}$ be the laminar family of minimum cuts in $H$, guaranteed by Lemma 2.3. Recall that for all $1 \leq i \leq r$, $|E_H(S_i, V \setminus S_i)| \leq k-1$. We need the following lemma.

**Lemma 4.2** *We can efficiently find a collection $\mathcal{P}$ of mutually disjoint vertex subsets, such that:*

- *For each $U \in \mathcal{P}$, $D(U) \leq r$;*
- *For each $U \in \mathcal{P}$, $|E_H(U,\overline{U})| \leq 2(k-1)$, and*
- $\sum_{U \in \mathcal{P}} D(U,\overline{U}) \geq \frac{r}{8\log r}$ .

**Proof:**
We will define each set $U \in \mathcal{P}$ to be either some set $S \in \mathcal{S}$, or a difference of two sets, $U = S \setminus S'$, for $S,S' \in \mathcal{S}$. Since for each set $S \in \mathcal{S}$, $D(S) \leq r$, this will ensure the first condition. Since $|E_H(U,\overline{U})| \leq |E_H(S,\bar{S})| + |E_H(S',\bar{S}')| \leq 2(k-1)$, this will also ensure the second condition.

We now turn to define the sets $U \in \mathcal{P}$ so that the third condition is also satisfied. For simplicity, if collection $\mathcal{S}$ contains identical sets, we discard them, keeping at most one copy of each set in $\mathcal{S}$. Recall that for each set $S \in \mathcal{S}$, $D(S,\bar{S})$ is the set of all demand pairs $(s_j,t_j)$ with $|\{s_j,t_j\} \cap S| = 1$. Let $D'(S,\bar{S})$ be the union of $D(S',\bar{S}')$ for all sets $S' \in \mathcal{S}$ where $S' \subset S$, and let $q(S) = |D(S,\bar{S}) \setminus D'(S,\bar{S})|$

We partition the family $\mathcal{S}$ into subsets $\mathcal{S}_x$, for $1 \leq x \leq \lfloor \log_2 r \rfloor + 1$, as follows: Collection $\mathcal{S}_x$ contains all sets $S \in \mathcal{S}$ with $2^{x-1} \leq q(S) < 2^x$. Since $\sum_{S\in\mathcal{S}} q(S) = r$, there is at least one index $x^*$, for which $\sum_{S\in\mathcal{S}_{x^*}} q(S) \geq \frac{r}{2\log r}$. Fix any such index $x^*$.

Consider the decomposition forest $\mathcal{F}$ for the sets in $\mathcal{S}_{x^*}$. The nodes of the forest are the sets in $\mathcal{S}_{x^*}$, and for a pair $S,S' \in \mathcal{S}_{x^*}$, set $S$ is the parent of $S'$ iff $S' \subset S$, and there is no other set $S'' \in \mathcal{S}_{x^*}$ with $S' \subset S'' \subset S$. Let $\mathcal{S}' \subseteq \mathcal{S}_{x^*}$ be the collection of sets that have at most one child in this forest. We are now ready to define the collection $\mathcal{P}$ of vertex subsets. If $S \in \mathcal{S}'$ is a leaf in $\mathcal{F}$, then we add $S$ to $\mathcal{P}$. Otherwise, if $S$ is a non-leaf set in $\mathcal{S}'$, and $S'$ is the unique child of $S$ in $\mathcal{F}$, then we add $S \setminus S'$ to $\mathcal{P}$.

It now only remains to prove that $\sum_{U\in\mathcal{P}} D(U,\overline{U}) \geq \frac{r}{8\log r}$. In order to do so, observe that $|\mathcal{S}'| \geq |\mathcal{S}_{x^*}|/2$, and recall that for each $S \in \mathcal{S}_{x^*}$, $2^{x^*-1} \leq q(S) < 2^{x^*}$. Therefore, $\sum_{U\in\mathcal{P}} D(U,\overline{U}) \geq \sum_{S\in\mathcal{S}'} q(S) \geq \frac{|\mathcal{S}_{x^*}|}{2} \cdot 2^{x^*-1} = |\mathcal{S}_{x^*}| \cdot 2^{x^*-2}$. On the other hand, $\sum_{S\in\mathcal{S}_{x^*}} q(S) \geq \frac{r}{2\log r}$, and so $|\mathcal{S}_{x^*}| \geq \frac{r}{2^{x^*+1}\log r}$. We conclude that $\sum_{U\in\mathcal{P}} D(U,\overline{U}) \geq \frac{r}{8\log r}$. □

Let $\mathcal{P}$ be the set family computed by Lemma 4.2. Clearly, for each $U \in \mathcal{P}$,

$$w^{(2k-1)}(U, \overline{U}) = \tilde{\Phi}^{(2k-1)}(U) \cdot D(U, \overline{U}) \tag{2}$$
$$\geq \tilde{\Phi}^{(2k-1)}(G) \cdot D(U, \overline{U}).$$

On the other hand, since $|E_H(U, \overline{U})| \leq 2k - 2$, $w(E^* \cap E_G(U, \overline{U})) \geq w^{(2k-1)}(U, \overline{U})$ must hold. Therefore,

$$\sum_{U \in \mathcal{P}} w^{(2k-1)}(U, \overline{U}) \leq \sum_{U \in \mathcal{P}} w(E^* \cap E_G(U, \overline{U})) \leq 2\mathsf{OPT}. \tag{3}$$

Combining Equations (2) and (3), we get that:

$$2\mathsf{OPT} \geq \sum_{U \in \mathcal{P}} w^{(2k-1)}(U, \overline{U}) \geq \tilde{\Phi}^{(2k-1)}(G) \sum_{U \in \mathcal{P}} D(U, \overline{U})$$
$$\geq \tilde{\Phi}^{(2k-1)}(G) \cdot \frac{r}{8 \log r}$$

Therefore, $\mathsf{OPT} = \Omega\left(\frac{r}{\log r}\right) \tilde{\Phi}^{(2k-1)}(G)$. □

In order to bound the cost $w(E')$ of the solution, we note that $D(S, \bar{S}) \subseteq D_0$, and so

$$w(E_0) = w^{(2k-1)}(S, \bar{S}) = \tilde{\Phi}^{(2k-1)}(S) \cdot D(S, \bar{S})$$
$$\leq \alpha_{\mathrm{ALN}}(r) \cdot \tilde{\Phi}^{(2k-1)}(G) \cdot |D_0|$$
$$= O(\alpha_{\mathrm{ALN}}(r) \log r) \cdot \frac{|D_0|}{r} \cdot \mathsf{OPT}.$$

We can now use Theorem 2.5 with $\alpha = O(\alpha_{\mathrm{ALN}}(r) \log r)$ to conclude that $w(E') = O(\alpha_{\mathrm{ALN}}(r) \log^2 r) = O(\log^{2.5} r \log \log r)$.

## 4.2 A polynomial-time bi-criteria approximation algorithm

In this section, we extend the algorithm from Section 4.1 to handle higher values of $k$ in polynomial time. Notice that the bottleneck in the algorithm from Section 4.1 is computing the approximate multi-route sparsest cut via Theorem 2.1, which is done in time $n^{O(k)}$. We use Theorem 2.2 instead, that gives an efficient algorithm for computing the $k$-route sparsest cut, albeit with somewhat weaker guarantees.

Our algorithm is identical to the algorithm in Figure 2, except for the following changes. First, in step 2, we use Theorem 2.2 to find an approximate $(2k - 1)$-route non-uniform sparsest cut $S$. That is, $\tilde{\Phi}^{(k')}(S) = O(\log r)\tilde{\Phi}^{(2k-1)}(G)$, where $k' = C(2k-1)\log r$, and $C$ is the constant from Theorem 2.2. Note that $\tilde{\Phi}^{(C \log r(2k-1))}(G) \leq \tilde{\Phi}^{(2k-1)}(G) \leq O\left(\frac{\log r}{r}\right) \mathsf{OPT}$ by Theorem 4.1. Therefore, $w(E_0) \leq \tilde{\Phi}^{(C \log r(2k-1))}(S) \cdot D(S, \bar{S}) \leq O(\log r) \tilde{\Phi}^{(C \log r(2k-1))}(G) \cdot D_0 \leq O\left(\frac{\log^2 r}{r}\right) \mathsf{OPT} \cdot D_0$. Using Theorem 2.5 with $\alpha = O(\log^2 r)$, we get that the algorithm returns a bi-criteria $(O(\log r), O(\log^3 r))$-approximate solution to the problem.

## 5 Vertex Connectivity

In this section, we extend our approximation algorithms for EC-kRC to handle vertex-connectivity and prove Theorem 1.3. We start by extending some of our technical definitions and results to the vertex-connectivity setting.

Let $(s,t)$ be any pair of vertices in graph $G$, and let $\Delta \subseteq V$ be any subset of vertices. We say that $\Delta$ is a *separator* for $s$ and $t$, or that $\Delta$ separates $s$ and $t$, iff $s, t \notin \Delta$, and $s$ and $t$ belong to two distinct connected components of $V \setminus \Delta$. We say that $\Delta$ is a minimum cost separator for $(s,t)$ iff for each subset $\Delta'$ separating $s$ from $t$, $|\Delta| \leq |\Delta'|$. Given any pair $S, T \subseteq V$ of vertex subsets, let $E(S, T)$ be the set of edges with one endpoint in $S$ and the other endpoint in $T$. Similarly, we say that $\Delta$ separates $S$ from $T$ iff $S \cap \Delta = \emptyset$, $T \cap \Delta = \emptyset$, and $E(S, T) = \emptyset$. Notice that in general $G \setminus \Delta$ may contain more than two connected components. A *vertex cut* in graph $G = (V, E)$ is a tri-partition $(S, \Delta, T)$ of $V$, where $E(S, T) = \emptyset$. For any subset $\Delta \subseteq V$, we will sometimes refer to $|\Delta|$ as the cost of $\Delta$.

We start with the following lemma, which is an analogue of Lemma 2.3 for vertex cuts. For technical reasons, it is more convenient to state it for graphs with costs on vertices. Given a graph $G = (V, E)$ with costs $c_v$ on vertices $v \in V$, a cost of a subset $\Delta \subseteq V$ of vertices is $\sum_{v \in \Delta} c_v$.

**Lemma 5.1** *Suppose we are given a graph $G = (V, E)$ with costs $c_v$ on vertices $v \in V$, and a set $\{(s_1, t_1), (s_2, t_2), \cdots, (s_r, t_r)\}$ of $r$ demand pairs. Let $\mathcal{T}$ be the set of all vertices participating in the demand pairs. Suppose additionally that for every demand pair $(s_i, t_i)$, for every minimum-cost separator $\Delta$ for $(s_i, t_i)$, $\Delta \cap \mathcal{T} = \emptyset$. Then there exists a family of vertex cuts $(S_i, \Delta_i, T_i)$ such that:*

1. *For every $i \in \{1, \cdots, r\}$, $\Delta_i$ is a minimum cost separator for $(s_i, t_i)$ (note that $s_i$ may belong either to $S_i$ or $T_i$); and*

2. *Sets $\{S_i\}_{i=1}^r$ form a laminar family.*

**Proof:**

We start by considering a "non-degenerate" case, where every subset $S \subseteq V$ of vertices has a distinct cost. Fix some $1 \leq i \leq r$. Let $\Delta_i$ be the unique minimum-cost separator for $(s_i, t_i)$. Consider the connected components of $G \setminus \Delta_i$ that contain $s_i$ and $t_i$. Let $S_i$ be the component of the smaller cost. We then set $T_i = V \setminus (S_i \cup \Delta_i)$. This finishes the definition of the cuts $(S_i, \Delta_i, T_i)$. Clearly, these cuts satisfy property 1.

We claim that sets $\{S_i\}_{i=1}^r$ form a laminar family. Assume for contradiction that this is not the case, so there are two sets $S_i, S_j$ whose intersection is non-empty, but neither of them is a subset of the other. Assume w.l.o.g. that these two sets are $S_1$ and $S_2$. Notice that $c(\Delta_1) \neq c(\Delta_2)$ must hold: otherwise, if $c(\Delta_1) = c(\Delta_2)$, then $\Delta_1 = \Delta_2$ must hold, and thus $S_1$ and $S_2$ are some connected components of $G \setminus \Delta_1$. But in that case, either $S_1 = S_2$ or $S_1 \cap S_2 = \emptyset$ must hold, a contradiction. Without loss of generality, we assume that $c(\Delta_1) > c(\Delta_2)$. Since $\Delta_1$ is the minimum cost separator for $s_1$ and $t_1$, and $c(\Delta_2) < c(\Delta_1)$, set $\Delta_2$ cannot be a separator for $s_1$ and $t_1$, and so both these vertices must either belong to $S_2$ or to $T_2$.

Let $X_2 \in \{S_2, T_2\}$ be the set that contains neither $s_1$ nor $t_1$, and let $Y_2$ be the other set (recall that by our assumption sets $\Delta_1$ and $\Delta_2$ contain no terminals). Recall that $X_2$ contains either $s_2$ or $t_2$ but not both of them. Let us assume, without loss of generality, that $s_2 \in X_2$ and $t_2 \in Y_2$. Let $X_1 \in \{S_1, T_1\}$ be set containing $s_2$, and let $Y_1$ be the other set. Assume without loss of generality that $s_1 \in S_1$ and $t_1 \in T_1$. Figure 3 shows which terminals lie in which sets.

|        | $X_2$ | $\Delta_2$ | $Y_2$      |
|--------|-------|------------|------------|
| $X_1$  | $s_2$ |            | $s_1, (t_2)$ |
| $\Delta_1$ |       |            |            |
| $Y_1$  |       |            | $t_1, (t_2)$ |

Figure 3: The figure shows the intersections of the sets $S_1, \Delta_1, T_1$ with the sets $S_2, \Delta_2, T_2$. There are edges only between sets located in horizontally, vertically, or diagonally adjacent cells. The figure also shows how terminals $s_1$, $t_1$, $s_2$, and $t_2$ are distributed among the sets, with the two possible locations of $t_2$ appearing in parentheses.

We need the following claim.

**Claim 5.1**
$$Y_1 \cap \Delta_2 = X_2 \cap \Delta_1 = \varnothing$$

.

|        | $X_2$ | $\Delta_2$ | $Y_2$      |
|--------|-------|------------|------------|
| $X_1$  | $s_2$ |            | $s_1, (t_2)$ |
| $\Delta_1$ |       |            |            |
| $Y_1$  |       |            | $t_1, (t_2)$ |

Set $A$

|        | $X_2$ | $\Delta_2$ | $Y_2$      |
|--------|-------|------------|------------|
| $X_1$  | $s_2$ |            | $s_1, (t_2)$ |
| $\Delta_1$ |       |            |            |
| $Y_1$  |       |            | $t_1, (t_2)$ |

Set $B$

|        | $X_2$ | $\Delta_2$ | $Y_2$      |
|--------|-------|------------|------------|
| $X_1$  | $s_2$ |            | $s_1, (t_2)$ |
| $\Delta_1$ | $\varnothing$ |            |            |
| $Y_1$  |       | $\varnothing$ | $t_1, (t_2)$ |

Sets $Y_1 \cap \Delta_2$ and $X_2 \cap \Delta_1$

Figure 4: Illustration for Claim 5.1

**Proof:** Let $A = (Y_1 \cap \Delta_2) \cup (Y_2 \cap \Delta_1) \cup (\Delta_1 \cap \Delta_2)$, and let $B = (X_1 \cap \Delta_2) \cup (X_2 \cap \Delta_1) \cup (\Delta_1 \cap \Delta_2)$ (see Figure 4). Notice that $A$ is a separator for $s_1$ and $t_1$, and $B$ is a separator for $s_2$ and $t_2$.

By our definition of cuts $(S_i, \Delta_i, T_i)$:

- either $c(\Delta_1) < c(A)$ or $\Delta_1 = A$; and
- either $c(\Delta_2) < c(B)$ or $\Delta_2 = B$.

However, $c(A) + c(B) = c(\Delta_1) + c(\Delta_2)$. Therefore, $\Delta_1 = A$ and $\Delta_2 = B$ must hold, and so $Y_1 \cap \Delta_2 = X_2 \cap \Delta_1 = \varnothing$. □

|       | $X_2$            | $\Delta_2$ | $Y_2$            |
|-------|------------------|------------|------------------|
| $X_1$ | $\mathbf{X_1}$   |            |                  |
| $\Delta_1$ | $\varnothing$ |         | $\Delta_1$       |
| $Y_1$ | $\mathbf{Y_1 \cap X_2}$ | $\varnothing$ | $\mathbf{Y_1 \cap Y_2}$ |

Thus sets $Y_1 \cap X_2$, $Y_1 \cap Y_2$ and $X_1$ are all disconnected in $G \setminus \Delta_1$. That is, $\Delta_1$ is a separator for each pair of these sets. We claim that $S_1 = X_1$ must hold. Indeed, since neither $s_1$ nor $t_1$ lie in $Y_1 \cap X_2$, $S_1 \neq Y_1 \cap X_2$. It is also impossible that $S_1 = Y_1 \cap Y_2$, since then either $S_1 \subseteq S_2$, or $S_1 \cap S_2 = \emptyset$, contradicting our initial assumption.

Similarly, sets $Y_1 \cap X_2$, $X_1 \cap X_2$, and $Y_2$ are disconnected in $G \setminus \Delta_2$.

|       | $X_2$            | $\Delta_2$ | $Y_2$ |
|-------|------------------|------------|-------|
| $X_1$ | $\mathbf{X_1 \cap X_2}$ | $\Delta_2$ |       |
| $\Delta_1$ | $\varnothing$ |       | $\mathbf{Y_2}$ |
| $Y_1$ | $\mathbf{Y_1 \cap X_2}$ | $\varnothing$ |  |

We claim that $S_2 = Y_2$ must hold. Indeed, $Y_1 \cap X_2$ does not contain $s_2$ or $t_2$, so $S_2 \neq Y_1 \cap X_2$. It is also impossible that $S_2 = X_1 \cap X_2$, since then $S_2 \subseteq X_1 \subseteq S_1$ must hold, contradicting our initial assumption. To summarize, we have shown that $S_1 = X_1$, and $S_2 = Y_2$ must hold. But then, by the definition of the sets $S_i$, $c(Y_1 \cap Y_2) > c(S_1) = c(X_1)$, and $c(X_1 \cap X_2) > c(S_2) = c(Y_2)$. Therefore, $c(Y_1 \cap Y_2) + c(X_1 \cap X_2) > c(X_1) + c(Y_2)$, which is impossible.

Finally, we show that we can perturb the costs of the vertices so that all subsets have different costs. Let

$$\delta = \min_{\substack{A,B \subset V: \\ c(A)-c(B)>0}} c(A) - c(B).$$

We assign a new cost $\tilde{c}_u$ to every vertex $u$ uniformly at random from the interval $[c_u, c_u + \delta/(2|V|)]$. Note that with probability 1, the costs of every two distinct vertex subsets will be different. We find a family of vertex cuts $(S_i, \Delta_i, T_i)$ w.r.t. the costs $\tilde{c}_v$. We verify that $\Delta_i$ is a minimum cost separator for $s_i$ and $t_i$ with respect to the original costs $c_v$. Assume for contradiction that there is a separator $\Delta$ for $s_i$ and $t_i$, with $c(\Delta) < c(\Delta_i)$.

Then

$$\tilde{c}(\Delta_i) \geq c(\Delta_i) \geq c(\Delta) + \delta > c(\Delta) + \frac{\delta}{2|V|} \cdot |\Delta| \geq \tilde{c}(\Delta),$$

which contradicts to the fact that $\Delta_i$ is the minimum cost separator for $s_i$ and $t_i$ w.r.t. costs $\tilde{c}$. □

### Proof of Theorem 1.3

In this section, we prove Theorem 1.3, by showing a $(2, O(dk \log^{5/2} r \log \log r))$ bi-criteria approximation algorithm VC-kRC, where $d$ is the maximum number of demand pairs in which any terminal participates. The running time of the algorithm is $n^{O(k)}$.

We start by extending the definition of the $k$-route sparsest cut to the vertex connectivity version. Given two disjoint subsets $S, T$ of vertices, let $D(S, T)$ be the set of all demand pairs $(s_i, t_i)$ with exactly one of the vertices $s_i, t_i$ lying in $S$, and the other one lying in $T$. Given any pair $(S, \Delta)$ of disjoint subsets, let $\Upsilon^{(\Delta)}(S) = \sum_{e \in E(S, V \setminus (S \cup \Delta))} w_e$, where $E(S, V \setminus (S \cup \Delta))$ is the subset of all edges with one endpoint in $S$ and the other endpoint in $V \setminus (S \cup \Delta)$.

The $k$-route vertex sparsity of the set $S$ is then defined to be:

$$\tilde{\Psi}^{(k)}(S) = \min_{\substack{\Delta \subseteq V \setminus S: \\ |\Delta| \leq k-1}} \left\{ \frac{\Upsilon^{(\Delta)}(S)}{|D(S, V \setminus (S \cup \Delta))|} \right\},$$

and the $k$-route vertex sparsity of the graph $G$ is:

$$\tilde{\Psi}^{(k)}(G) = \min_{S \subset V} \left\{ \tilde{\Psi}^{(k)}(S) \right\}$$

It is easy to see that, similarly to the edge version of $k$-route sparsest cut, the $k$-route vertex sparsest cut can be approximated in time $n^{O(k)}$ to within a factor of $\alpha_{\text{ALN}}(r)$, as we show in the next theorem.

**Theorem 5.2** *There is an algorithm that finds, in time $n^{O(k)}$, disjoint subsets $S, \Delta \subset V$ of vertices, with $|\Delta| \leq k - 1$ such that*

$$\Upsilon^{(\Delta)}(S) \leq \alpha_{\text{ALN}}(r) \cdot \tilde{\Psi}^{(k)}(G) \cdot |D(S, V \setminus (S \cup \Delta))|.$$

**Proof:** For every subset $\Delta \subset V$ of at most $k - 1$ vertices, we use the algorithm $\mathcal{A}_{\text{ALN}}$ to find an approximate sparsest cut in the graph $G \setminus \Delta$, and output the sparsest cut among all such cuts. □

Our algorithm for VC-kRC is very similar to the algorithm for EC-kRC from Section 4. The only difference is that we use Theorem 5.2 to find an approximate $k$-route vertex sparsest cut. The algorithm is summarized in Figure 5.

It is easy to verify that if $E'$ is the solution computed by the algorithm, then for each demand pair $(s_i, t_i)$ there are at most $(2k - 1)$ vertex-disjoint paths connecting them in $G \setminus E'$. This is since the algorithm only removes a demand pair $(s_i, t_i)$ when the terminals $s_i$ and $t_i$ are no longer $(2k - 1)$-vertex connected, and it terminates, since it removes at least one demand pair in each iteration.

In order to analyze the performance of the algorithm, we use the following theorem, that relates the value $\tilde{\Psi}^{(k)}(G)$ of the $k$-route vertex sparsest cut in graph $G$ to the value OPT of the optimal solution to VC − kRC.

**Theorem 5.3**

$$\tilde{\Psi}^{(2k-1)}(G) \leq O\left(\frac{dk \log r}{r}\right) \cdot \text{OPT}.$$

**Proof:** Let $H = G \setminus E^*$. The proof roughly follows the proof of Theorem 3.1, except that we need one additional step, that is summarized in the following lemma.

**Lemma 5.4** *There exists a subset $D' \subseteq D$ of $r' = \Omega(r/(dk))$ demand pairs, and a collection of vertex cuts $\{(S_i, \Delta_i, T_i)\}_{(s_i, t_i) \in D'}$, such that:*

---

**Input**: A weighted graph $G(V, E)$ with a set $D = \{(s_i, t_i)\}_{1 \leq i \leq r}$ of demand pairs, and edge weights $\{w_e\}_{e \in E}$.

**Output**: A subset $E'$ of edges, such that no demand pair $s_i$ and $t_i$ is $(2k - 1)$-vertex connected in $G \setminus E'$.

1. If $r = 0$ return $E' = \emptyset$.

2. Find sets $U$ and $\Delta$ with $|\Delta| \leq 2k-1$ and $\Upsilon^{(\Delta)}(U) \leq \alpha_{\text{ALN}}(r) \cdot \tilde{\Psi}^{(2k-1)}(G) \cdot |D(U, V \setminus (U \cup \Delta))|$ using Theorem 5.2.

3. Let $E_0 = E(U, V \setminus (U \cup \Delta))$, and let $G' = G \setminus E_0$.

4. Let $D_0$ be the set of all demand pairs $(s_i, t_i)$ that are no longer $(2k - 1)$-vertex connected in $G'$.

5. Solve the problem recursively on $G'$ with the set $D \setminus D_0$ of demand pairs to obtain a solution $E_1$.

6. Return $E' = E_0 \cup E_1$.

---

Figure 5: Bi-criteria approximation algorithm for VC-kRC in time $n^{O(k)}$.

- *For all $(s_i, t_i) \in D'$, $\Delta_i$ is a separator for $(s_i, t_i)$ in $H$, $|\Delta_i| < k$, and $\Delta_i \cap T' = \emptyset$, where $T'$ is the set of all terminals participating in demand pairs in $D'$.*

- $\{S_i\}_{(s_i, t_i) \in D'}$ *is a laminar family of vertex subsets.*

**Proof:** For each $1 \leq i \leq r$, let $\Delta'_i$ be a minimum vertex separator for $s_i$ and $t_i$ in $H$. Since $s_i$ and $t_i$ are not $k$-vertex connected in $H$, $|\Delta'_i| < k$. We construct an auxiliary graph $Z$, whose vertex set is $[r]$, that is, each vertex $i$ of $Z$ represents the demand pair $(s_i, t_i)$. We say that a demand $i$ *blocks* another demand $j$ iff $\Delta'_i$ contains either $s_j$ or $t_j$ (or both). We connect $i$ and $j$ with an edge in $Z$ iff one of them blocks the other. Since $|\Delta'_i| \leq k - 1$ and each vertex in $\Delta'_i$ participates in at most $d$ demand pairs, demand $i$ blocks at most $d(k - 1)$ demands. Therefore, there are at most $d(k-1)r$ edges in $Z$. By Turan's theorem, there is an independent set $I$ of size $\Omega(r/(dk))$ in $Z$. Let $r' = |I|$, and let $D' = \{(s_i, t_i) \mid i \in I\}$.

Next, we apply Lemma 5.1 to graph $G$ with the set $D'$ of demand pairs, where we define the cost $c_u$ of every vertex $u$ as follows: $c_u = |V|$ if $u = s_i$ or $u = t_i$ for some $(s_i, t_i) \in D'$, and $c_u = 1$ otherwise. Since demand pairs in $D'$ do not block each other, the minimum *cost* vertex cut for each of them has cost at most $k - 1 < |V|$. Let $\{(S_i, \Delta_i, T_i)\}_{(s_i, t_i) \in D'}$ be the collection of cuts returned by Lemma 5.1. It is easy to see that these cuts satisfy the conditions of the lemma. □

We use Lemma 5.4 to find a subset $D'$ of demand pairs and vertex cuts $(S_i, \Delta_i, T_i)$. We assume w.l.o.g. that $D' = \{(s_1, t_1), \ldots, (s_{r'}, t_{r'})\}$. Now we need the following counterpart of Lemma 4.2.

**Lemma 5.5** *There is a family $\mathcal{P} = \{U_1, \ldots, U_p\}$ of disjoint vertex subsets, and a collection $\{(U_j, \Lambda_j, R_j)\}_{j=1}^{p}$ of vertex cuts in graph $H$, such that:*

- *for each $1 \leq j \leq p$, $|\Lambda_j| < 2k - 1$,*

- $\sum_{j=1}^{p} |D(U_j, R_j)| \geq \frac{r'}{8 \log r'}$.

**Proof:** The proof closely follows the proof of Lemma 4.2. Let $\mathcal{S} = \{S_1, \ldots, S_{r'}\}$ be the family of vertex subsets from Lemma 5.4, and assume that the vertex cut corresponding to set $S_i \in \mathcal{S}$ is $(S_i, \Delta_i, T_i)$. Family $\mathcal{P}$ will contain two type of vertex subsets. Subset $U_j$ is a subset of the first type iff $U_j = S_i$ for some $S_i \in \mathcal{S}$. In this case, we set $\Lambda_j = \Delta_i$, and the corresponding cut $(U_j, \Lambda_j, R_j) = (S_i, \Delta_i, T_i)$. It is easy to see that the first condition of the lemma will hold for vertex subsets of this type.

Subset $U_j$ of vertices is a subset of the second type iff $U_j = S_i \setminus (S_{i'} \cup \Delta_{i'})$ for some $S_i, S_{i'} \in \mathcal{S}$, where $S_{i'} \subset S_i$. In this case, we set $\Lambda_j = \Delta_i \cup \Delta_{i'}$, and $R_j = V \setminus (U_j \cup \Lambda_j)$. Notice that if $e = (u, v) \in E(H)$ has $u \in U_j$, $v \notin U_j$, then $v \in \Lambda_j$ must hold. Indeed, if $v \notin S_i$, then since $(S_i, \Delta_i, T_i)$ is a valid vertex cut, $v \in \Delta_i$ must hold. Otherwise, if $v \in S_i$, but $v \notin \Delta_{i'}$, then $v \in S_{i'}$ must hold, and since $(S_{i'}, \Delta_{i'}, T_{i'})$ is a valid vertex cut, $u \in \Delta_{i'}$ must hold, which is impossible. Therefore, $(U_j, \Lambda_j, R_j)$ is a valid vertex cut. Moreover, $|\Lambda_j| = |\Delta_i \cup \Delta_{i'}| \leq 2(k-1)$, and so the first condition of the lemma holds.

We now show how to define the family $\mathcal{P}$, so that the second condition of the lemma is satisfied as well. We assume w.l.o.g. that $\mathcal{S}$ does not contain two copies of the same set: otherwise, we simply remove copies of the same set, until just one copy remains in $\mathcal{S}$.

For every set $S_i \in \mathcal{S}$, let $D'(S_i) = \bigcup_{\substack{S_j \in \mathcal{S}: \\ S_j \subset S_i}} D(S_j, T_j) \cap D'$, and let $q(S_i) = |(D(S_i, T_i) \cap D') \setminus D'(S_i)|$. As before, we partition the set $\mathcal{S}$ as follows: for $x = 1, \ldots, \lfloor \log_2 r' \rfloor + 1$, let $\mathcal{S}_x = \{S_i \in \mathcal{S} \mid 2^{x-1} \leq q(S_i) < 2^x\}$.

Since $\sum_{x=1}^{\lfloor \log_2 r' \rfloor + 1} \sum_{S_i \in \mathcal{S}_x} q(S_i) = r'$, we can choose an index $x^*$, such that $\sum_{S_i \in \mathcal{S}_{x^*}} q(S_i) \geq \frac{r'}{2 \log r'}$. We say that a set $S_i$ is *good* if it belongs to $\mathcal{S}_{x^*}$. Consider the decomposition forest $\mathcal{F}$ for the good sets $S_i$: the nodes of the forest are the sets of $\mathcal{S}_{x^*}$, and $S_i$ is the parent of $S_j$ iff $S_j \subset S_i$, and there is no other set $S_\ell \in \mathcal{S}_{x^*}$ with $S_j \subset S_\ell \subset S_i$. Let $\mathcal{S}'$ be the subset of nodes of $\mathcal{F}$ with at most one child. Note that

$$|\mathcal{S}'| \geq |\mathcal{S}'_{x^*}|/2 \geq \frac{\sum_{S_i \in \mathcal{S}'_{x^*}} q(S_i)}{2 \cdot 2^{x^*}} \geq \frac{r'}{2^{x^*+1}(2 \log r')}.$$

On the other hand, since $q(S_i) \geq 2^{x^*-1}$ for $S_i \in \mathcal{S}'$,

$$\sum_{S_i \in \mathcal{S}'} q(S_i) \geq \frac{r'}{8 \log r'} \qquad (4)$$

For every set $S_i \in \mathcal{S}'$, we let $U_i = S_i$ and $\Lambda_i = \Delta_i$ if $S_i$ is a leaf of $\mathcal{F}$; we let $U_i = S_i \setminus (S_j \cup \Delta_j)$ and $\Lambda_i = \Delta_i \cup \Delta_j$ if $S_i$ has a unique child $S_j$ in $\mathcal{F}$. Let $R_i = V \setminus (U_i \cup \Lambda_i)$.

If set $U_i$ is of the first type, then $D(U_i, R_i) = D(S_i, T_i)$, and so $|D(U_i, R_i) \cap D'| \geq q(S_i)$. Otherwise, if $U_i = S_i \setminus (S_j \cup \Delta_j)$, then $D(U_i, R_i) \cap D'$ contains all demand pairs in $D(S_i, T_i) \cap D'$, except for pairs $(x, y)$ with $x \in S_j \cup \Delta_j$, $y \in T_i$. But since $\Delta_j$ does not contain any terminals participating in pairs in $D'$, $x \in S_j$, $y \in T_j$ and $(x, y) \in D(S_j, T_j)$ must hold. Therefore, $|D(U_i, R_i) \cap D'| \geq q(S_i)$, and so $\sum_{U_i \in \mathcal{P}} |D(U_i, R_i)| \geq \frac{r'}{8 \log r'}$. □

Consider the family $\mathcal{P} = \{U_1, \ldots, U_p\}$ and the corresponding cuts $(U_i, \Lambda_i, R_i)$ as in Lemma 5.5. Since all sets in $\mathcal{P}$ are mutually disjoint, and for each such set $U_i \in \mathcal{P}$, $|E_H(U_i, R_i)| \leq 2k-1$,

$$\sum_{j=1}^{p} \Upsilon^{(\Lambda_j)}(U_j) \leq 2\mathsf{OPT},$$

and so

$$\frac{\sum_{j=1}^{p} \Upsilon^{(\Lambda_j)}(U_j)}{\sum_{j=1}^{p} |D(U_j, R_j)|} \leq O\left(\frac{\log r'}{r'}\right) \cdot \mathsf{OPT}$$
$$\leq O\left(\frac{dk \log r}{r}\right) \cdot \mathsf{OPT}.$$

Therefore, there is an index $1 \leq j \leq p$, such that

$$\frac{\Upsilon^{(\Lambda_j)}(U_j)}{|D(U_j, R_j)|} \leq O\left(\frac{dk \log r}{r}\right) \cdot \mathsf{OPT}.$$

The left hand side of this inequality is at least $\tilde{\Psi}^{(2k-1)}(G)$ since $|\Lambda_j| \leq 2k - 2$. We conclude that $\tilde{\Psi}^{(2k-1)}(G) \leq O\left(\frac{dk \log r}{r}\right) \cdot \mathsf{OPT}$. □

In order to complete the proof of Theorem 1.3, observe that $w(E_0) = \Upsilon^{(\Delta)}(U)$, and by Theorem 5.3,

$$w(E_0) \leq \alpha_{\mathrm{ALN}}(r) \Psi^{(2k-1)}(G) |D(U, V \setminus (U \cup \Delta))|$$
$$\leq O\left(\frac{dk \log r}{r}\right) \cdot \alpha_{\mathrm{ALN}}(r) \cdot \mathsf{OPT} \cdot |D(U, V \setminus (U \cup \Delta))|.$$

Note that we remove all demand pairs in $D(U, V \setminus (U \cup \Delta))$ in step 4 of the algorithm. We can now use Theorem 2.5 with $\alpha = O(dk \log r \cdot \alpha_{\mathrm{ALN}}(r))$ to conclude that the cost of the solution returned by the algorithm is bounded by $O(dk \log^{5/2} r \log \log r) \cdot \mathsf{OPT}$.

# 6 Algorithms for 2-route cuts

In this section we prove Theorem 1.5. Since we prove in Section A that EC-kRC can be cast as a special case of VC-kRC, and the connectivity value $k$ remains unchanged in this reduction, it is enough to prove the theorem for VC-kRC, where $k = 2$. In the rest of this section we show an efficient $O(\log^{3/2} r)$-approximation algorithm for VC-kRC with $k = 2$.

Given a subset $S$ of vertices in graph $G$, the uniform vertex 2-route sparsity of $S$ is:

$$\Psi^{(2)}(S) = \min_{\substack{\Delta \subseteq V \setminus S: \\ |\Delta| \leq 1}} \left\{ \frac{\Upsilon^{(\Delta)}(S)}{\min\{D(S), D(V \setminus (S \cup \Delta))\}} \right\},$$

and the uniform vertex 2-route sparsity of the graph $G$ is:

$$\Psi^{(2)}(G) = \min_{S \subset V} \left\{ \Psi^{(2)}(S) \right\}$$

As before, we can efficiently approximate the uniform vertex 2-route sparsest cut in any graph, as shown in the next theorem.

**Theorem 6.1** *There is a polynomial time algorithm that finds disjoint subsets $S \subset V$ and $\Delta \subset V$ of vertices, with $|\Delta| \leq 1$ and $0 < D(S) \leq r$, such that*

$$\Upsilon^{(\Delta)}(S) \leq \alpha_{\text{ARV}}(r) \cdot \Psi^{(2)}(G) \cdot D(S).$$

**Proof:** For every subset $\Delta \subset V$ of size at most 1, we use the algorithm $\mathcal{A}_{\text{ARV}}$ to find the $\alpha_{\text{ARV}}(r)$-approximate uniform sparsest cut in graph $G \setminus \Delta$, and output the cut with the smallest sparsity. □

The approximation algorithm for VC-kRC with $k = 2$ is shown in Figure 6.

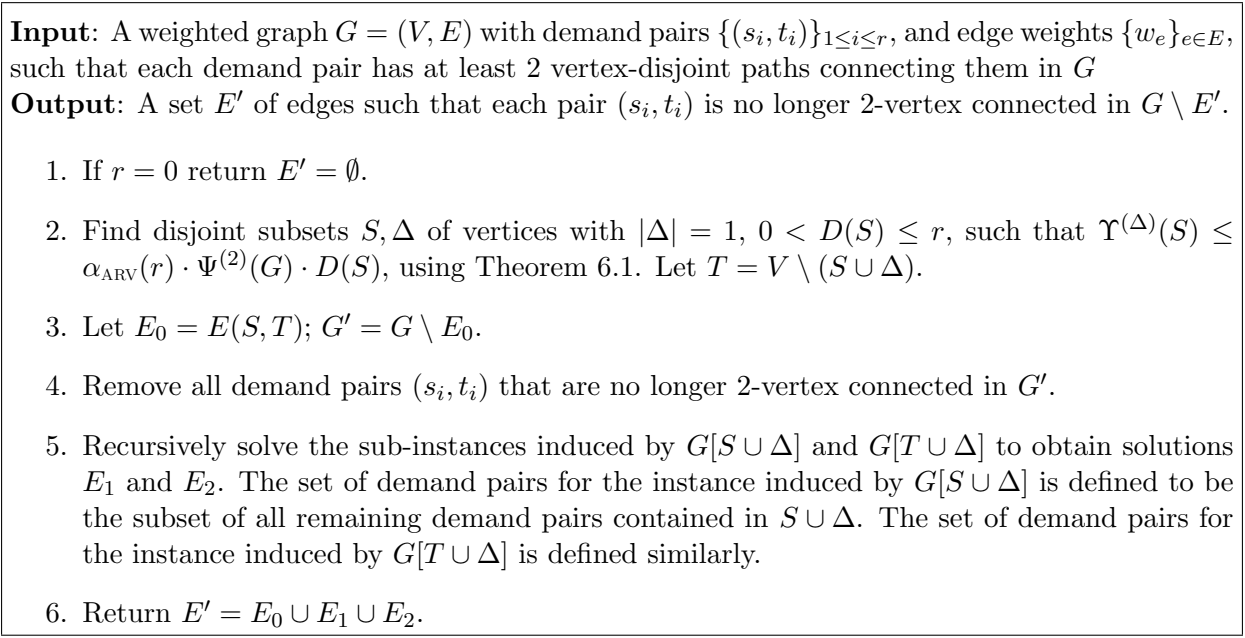

---

**Input**: A weighted graph $G = (V, E)$ with demand pairs $\{(s_i, t_i)\}_{1 \leq i \leq r}$, and edge weights $\{w_e\}_{e \in E}$, such that each demand pair has at least 2 vertex-disjoint paths connecting them in $G$.
**Output**: A set $E'$ of edges such that each pair $(s_i, t_i)$ is no longer 2-vertex connected in $G \setminus E'$.

1. If $r = 0$ return $E' = \emptyset$.

2. Find disjoint subsets $S, \Delta$ of vertices with $|\Delta| = 1$, $0 < D(S) \leq r$, such that $\Upsilon^{(\Delta)}(S) \leq \alpha_{\text{ARV}}(r) \cdot \Psi^{(2)}(G) \cdot D(S)$, using Theorem 6.1. Let $T = V \setminus (S \cup \Delta)$.

3. Let $E_0 = E(S, T)$; $G' = G \setminus E_0$.

4. Remove all demand pairs $(s_i, t_i)$ that are no longer 2-vertex connected in $G'$.

5. Recursively solve the sub-instances induced by $G[S \cup \Delta]$ and $G[T \cup \Delta]$ to obtain solutions $E_1$ and $E_2$. The set of demand pairs for the instance induced by $G[S \cup \Delta]$ is defined to be the subset of all remaining demand pairs contained in $S \cup \Delta$. The set of demand pairs for the instance induced by $G[T \cup \Delta]$ is defined similarly.

6. Return $E' = E_0 \cup E_1 \cup E_2$.

---

Figure 6: Approximation algorithm for VC-kRC, $k = 2$ (weighted case).

In order to analyze the algorithm, we start by showing that it is guaranteed to produce a feasible solution.

**Claim 6.1** *The algorithm outputs a feasible solution to the problem.*

**Proof:** The proof is by induction on the number of vertices in $G$. Assume that the algorithm outputs a feasible solution for all graphs containing fewer than $n$ vertices, and consider a graph $G$ containing $n$ vertices. Let $(s_i, t_i)$ be any demand pair, and assume for contradiction that there are at least two vertex-disjoint simple paths $P_1, P_2$ connecting $s_i$ to $t_i$ in $G \setminus E'$. Observe first that either $s_i, t_i \in S \cup \Delta$ or $s_i, t_i \in T \cup \Delta$ must hold. Otherwise, one of the two vertices must belong to $S$ and the other to $T$. But $\Delta$ is a separator for $S$ and $T$ in graph $G'$, and since $|\Delta| = 1$, the paths $P_1$ and $P_2$ cannot be vertex-disjoint. Assume w.l.o.g. that $s_i, t_i \in S \cup \Delta$. By the induction hypothesis, $E_1$ is a feasible solution to the instance induced by $G[S \cup \Delta]$, and in particular $G[S \cup \Delta] \setminus E_1$ cannot contain two vertex-disjoint paths connecting $s_i$ to $t_i$. Therefore, at least one of the two paths, say $P_1$, must contain a vertex of $T$. But since $\Delta$ is a separator for $S$ and $T$, $|\Delta| = 1$, and both $s_i, t_i \notin T$, path $P_1$ cannot be a simple path, a contradiction. □

It now remains to bound the cost of the solution produced by the algorithm. As before, we do so by relating the value of the 2-route vertex sparsest cut to the value OPT of the optimal solution to the VC-kRC problem.

**Theorem 6.2** *Suppose that we are given an undirected graph $G = (V, E)$ with edge weights $w_e$, and $r$ demand pairs $(s_1, t_1), \ldots, (s_r, t_r)$. Let OPT be cost of the optimal solution to the corresponding VC-kRC problem instance, and assume that $k = 2$. Then*

$$\Psi^{(2)}(G) \leq \frac{4\mathsf{OPT}}{r}.$$

**Proof:** We will assume that $G$ is connected: if $G$ is not connected and all terminals lie in one connected component then we just let $G$ to be this connected component; otherwise, if some terminals lie in one connected component and others lie in another connected component then $\Psi^{(2)}(G) = 0$ and we are done. Consider the graph $H = G \setminus E^*$. By the optimality of $E^*$, graph $H$ is connected. Since $E^*$ is a solution to the VC-kRC problem with $k = 2$, for every demand pair $(s_i, t_i)$, there is a one-vertex subset $\Delta_i$ of vertices separating $s_i$ from $t_i$ in $H$.

Consider the block decomposition of $H$. Recall that a block $B$ of $H$ is a maximal 2-vertex connected subgraph of $H$. Every pair $(B_1, B_2)$ of distinct blocks share either no vertices or only one vertex, and in the latter case, this vertex is a cut vertex. Let $\mathcal{B}$ be the set of all blocks, and let $S$ be the set of all cut vertices of $H$. The block tree $B(H)$ of $H$ is a bipartite graph, with parts $\mathcal{B}$ and $S$, in which a block $B$ is connected to a cut vertex $u$ iff $u$ lies in $B$. We assign costs $c(u)$ to each node $u$ of the tree $B(H)$ as follows. If $u$ is a cut vertex of $H$, let $c(u) = D_u$ (the number of demand pairs in which $u$ participates); if $B$ is a block, then $c(B) = D(V(B) \setminus S)$. Then each vertex $u \in S$ contributes $D_u$ to the cost of $u$ in $B(H)$; each vertex $u \in V \setminus S$ contributes $D_u$ to the cost of the block that contains $u$. Thus the total cost of all vertices in $B(H)$ is exactly $2r$. Note that the cost $c(u)$ of a cut vertex $u$ is at most $r$ (since there are $r$ demand pairs); the cost $c(u)$ of a block $B$ is also at most $r$ since for every $i \in [r]$ the block $B$ contains at most one of the terminals $s_i$ and $t_i$.

Assume first that we can find a block $B$ in the tree $B(H)$, such that, if $u_1, \ldots, u_p$ are the neighbors of $B$ in the tree $B(H)$ (recall that they must be the cut points of $H$ that lie in $B$), and for each $1 \leq i \leq p$, $\mathcal{T}_i$ is the subtree of $B(H)$ rooted at $u_i$, then $c(\mathcal{T}_i) \leq 5r/4$.

For each $1 \leq i \leq p$, let $S_i$ be the union of all blocks whose corresponding nodes lie in the tree $\mathcal{T}_i$, excluding the vertex $u_i$, that is, $S_i = \left(\bigcup_{B' \in \mathcal{T}_i} V(B')\right) \setminus \{u_i\}$, and let $T_i = V \setminus (S_i \cup \{u_i\})$.

Note that $u_i$ is a separator vertex for $S_i$ and $T_i$ in $H$. By our choice of $x$, $D(S_i) \leq 5r/4$, and $D(T_i) = 2r - D(S_i) - D(u_i) \geq 3r/4 \geq D(S_i)/2$ (since $D(S_i) + D(u_i) = c(\mathcal{T}_i) \leq 5r/4$). On the other hand, $\sum_{i=1}^{p} D(S_i) = 2r - D(B) \geq r$.

For each $1 \leq i \leq p$, there are no edges between $S_i$ and $V \setminus (S_i \cup \{u_i\})$ in $H$, thus we have

$$\begin{aligned} w(E^* \cap E_G(S_i, T_i)) &= w(E_G(S_i, T_i)) \\ &\geq \Psi^{(2)}(G) \cdot \min\{D(S_i), D(T_i)\} \\ &\geq \Psi^{(2)}(G) \cdot D(S_i)/2. \end{aligned}$$

Summing this inequality over all $1 \leq i \leq p$, we get

$$\begin{aligned} 2\mathsf{OPT} &\geq \sum_{i=1}^{p} w(E^* \cap E_G(S_i, T_i)) \geq \sum_{i=1}^{p} \Psi^{(2)}(G) \cdot D(S_i)/2 \\ &\geq \Psi^{(2)}(G) \cdot r/2. \end{aligned}$$

We conclude that $\Psi^{(2)}(G) \leq 4\mathsf{OPT}/r$.

We find a node $x$ in the tree $B(H)$ with the following property: if we root the tree $B(H)$ at $x$, then for each child $x'$ of $x$, the cost of the subtree rooted at $x'$ is at most $r$ (half of the cost of $B(H)$). In order to find such a node $x$, start with an arbitrary node $x_0 = x$, and root the tree $B(H)$ at $x$. As long as $x$ has a child $x'$, such that the cost of the sub-tree rooted at $x'$ is more than $r$, we set $x = x'$ and continue. Since during this process we always move down the tree rooted at $x_0$, it is guaranteed to terminate. It is easy to verify that the node $x$ at which the process terminates has the desired properties. If the node $x$ where this process terminates is a block, then from the above discussion, $\Psi^{(2)}(G) \leq 4\mathsf{OPT}/r$. Therefore, we assume that $x$ is a cut vertex. Let $\mathcal{T}_1, \ldots, \mathcal{T}_p$ be the set of the subtrees of $B(H)$ rooted at the children nodes of $x$, sorted by their cost, with $\mathcal{T}_1$ being the most expensive subtree and $\mathcal{T}_p$ the cheapest one.

Assume first that $c(\mathcal{T}_1) \leq r/2$. Note that $\sum_{i=1}^{p} c(\mathcal{T}_i) = 2r - c(x) \geq r$. Let $j^*$ be the largest index $j$ such that $\sum_{i=1}^{j} c(\mathcal{T}_i) \leq r/2$. By our choice of $x$, we have $1 \leq j^* < p$. Clearly, $\sum_{i=1}^{j^*} c(\mathcal{T}_i) \geq \frac{r}{4}$.

Let $S$ be the set of all vertices of $H$ contained in all blocks $B \in V(\mathcal{T}_i)$, for all $1 \leq i \leq j^*$ (excluding the vertex $x$), and let $T = V \setminus (S \cup \{x\})$. Then $D(T) = 2r - D(S) - D(x) \geq r/2$. Therefore, $\Psi^{(2)}(G) \leq \Psi^{(2)}(S) \leq \frac{\Upsilon^{(\{x\})}(S)}{\min\{D(S), D(T)\}} \leq \frac{\mathsf{OPT}}{r/4}$.

Assume now that $c(\mathcal{T}_1) > r/2$, but $c(\mathcal{T}_1) + c(x) \leq 7r/4$. Then $\sum_{j=2}^{p} c(\mathcal{T}_p) \geq r/4$. Setting $S$ to be the union of all blocks whose vertices lie in the trees $\mathcal{T}_j$ for $j \neq 1$ (excluding the vertex $x$), and setting $T = V \setminus (S \cup \{x\})$, we get that $D(S) \geq r/4$, $D(T) \geq r/2$, and we get that $\Psi^{(2)}(G) \leq \Psi^{(2)}(S) \leq \frac{\Upsilon^{(\{x\})}(S)}{\min\{D(S), D(T)\}} \leq \frac{\mathsf{OPT}}{r/4}$ as before.

Finally, if $c(\mathcal{T}_1) + c(x) > 7r/4$, let $B$ denote the child of $x$ that serves as the root of $\mathcal{T}_1$. Then $B$ has the property that every tree rooted at a neighbor of $B$ has a cost of at most $5r/4$, and as we have shown, $\Psi^{(2)}(G) \leq 4\mathsf{OPT}/r$ in this case. $\square$

Let $a$ be the number of demand pairs contained in $S \cup \Delta$ and $b$ be the number of demand pairs contained in $(V \setminus S) \cup \Delta$ in graph $G'$. From Theorem 6.2,

$$\begin{aligned} w(E_0) &\leq \alpha_{\mathrm{ARV}}(r) \cdot \Psi^{(2)}(G) \cdot D(S) \\ &\leq 4\alpha_{\mathrm{ARV}}(r)\mathsf{OPT}\min\{a, b\}/r \\ &\leq 4\alpha_{\mathrm{ARV}}(r)\mathsf{OPT}\min\{r - a, r - b\}/r \end{aligned}$$

Therefore, by setting $\alpha = 4\alpha_{\text{ARV}}(r)$, we get the same recurrence as in the proof of Theorem 2.4:

$$w(E') \leq w(E'_1) + w(E'_2) + 2\alpha \cdot \frac{\min\{r-a, r-b\} \, \mathsf{OPT}}{r}$$

Solving this recurrence as in Theorem 2.4, we get that $w(E') \leq O(\log^{3/2} r)\mathsf{OPT}$.

# 7 A factor $k^\epsilon$-hardness for $k$-VC-kRC

In this section we prove Theorem 1.4. We perform a reduction from the 3SAT(5) problem. In this problem we are given a 3SAT formula $\varphi$ on $n$ variables and $5n/3$ clauses. Each clause contains 3 distinct literals and each variable participates in exactly 5 different clauses. We say that $\varphi$ is a Yes-Instance if it is satisfiable. We say that $\varphi$ is a No-Instance with respect to some parameter $\epsilon$, iff no assignment satisfies more than $\epsilon$-fraction of clauses. The following well-known theorem follows from the PCP theorem [AS98, ALM$^+$98].

**Theorem 7.1** *There is a constant $\epsilon : 0 < \epsilon < 1$, such that it is NP-hard to distinguish between* Yes-Instances *and* No-Instances *(defined with respect to $\epsilon$) of the 3SAT(5) problem.*

We use the Raz verifier for 3SAT(5) with $\ell$ parallel repetitions. This is an interactive proof system, in which two provers try to convince the verifier that the input 3SAT(5) formula $\varphi$ is satisfiable. The verifier chooses, independently at random, $\ell$ clauses $C_1, \ldots, C_\ell$, and for each $i : 1 \leq i \leq \ell$, a variable $x_i$ participating in clause $C_i$ is chosen at random. The verifier then sends one query to each one of the two provers, while the query to the first prover consists of the indices of the variables $x_1 \ldots, x_\ell$, and the query to the second prover contains the indices of the clauses $C_1, \ldots, C_\ell$. The first prover returns an assignment to variables $x_1, \ldots, x_\ell$. The second prover is expected to return an assignment to all the variables in clauses $C_1, \ldots, C_\ell$, which must satisfy the clauses. Finally, the verifier checks that the answers of the two provers are consistent, i.e., for each $i : 1 \leq i \leq \ell$, the assignment to $x_i$, returned by the first prover, is identical to the assignment to $x_i$, obtained by projecting the assignment to the variables of $C_i$, returned by the second prover, onto $x_i$. (We assume that the answers sent by the second prover always satisfy the clauses appearing in its query). The following theorem is obtained by combining the PCP theorem with the parallel repetition theorem [Raz98].

**Theorem 7.2 ( [AS98, ALM$^+$98, Raz98])** *There exists a constant $\gamma > 0$, such that:*

- *If $\varphi$ is a* Yes-Instance*, then there is a strategy of the provers, for which the acceptance probability is $1$.*

- *If $\varphi$ is a* No-Instance*, then for any strategy of the provers, the acceptance probability is at most $2^{-\gamma \ell}$.*

We denote the set of all the random strings of the verifier by $R$, $|R| = (5n)^\ell$, and the sets of all the possible queries of the first and the second prover by $Q_1$ and $Q_2$ respectively, $|Q_1| = n^\ell$, $|Q_2| = (5n/3)^\ell$. For each query $q \in Q$, let $A(q)$ be the collection of all the possible answers to $q$ (if $q$ is a query of the second prover, then $A(q)$ only contains answers that satisfy all the clauses of the query). Let $A = 2^\ell$, $A' = 7^\ell$. Then for each $q \in Q_1$, $|A(q)| = A$, and for each $q' \in Q_2$, $|A(q')| = A'$. Given a random string $r \in R$, let $q_1(r), q_2(r)$ be the queries sent to the first and the second prover respectively, when the verifier chooses $r$. For each $q \in Q_1$, let $N(q) = \{q' \in Q_2 \mid \exists r \in R : q_1(r) = q, q_2(r) = q'\}$, and for each $q' \in Q_2$, let $N(q') = \{q \in Q_1 \mid \exists r \in R : q_1(r) = q, q_2(r) = q'\}$. Notice that for all $q \in Q_1$, $|N(q)| = 5^\ell$, and for all $q' \in Q_2$, $|N(q')| = 3^\ell$.

*Construction:* We now turn to describe our reduction. For each query $q \in Q_1$ of the first prover, for each answer $a \in A(q)$, we have an edge $e(q, a)$, whose endpoints are denoted by $v(q, a), u(q, a)$, and whose cost is $(5/3)^\ell$. We will think of $v(q, a)$ as the first endpoint of $e(q, a)$ and of $u(q, a)$ as its second endpoint, even though the graph is undirected. Similarly, for each query $q \in Q_2$ of the second prover, for each answer $a \in A(q)$, there is an edge $e(q, a) = (v(q, a), u(q, a))$, of cost 1. As before, $v(q, a)$ is called the first endpoint and $u(q, a)$ the second endpoint of $e(q, a)$. Let $E_0$ be the set of all resulting edges. For each $q \in Q$, let $V(q) = \{v(q, a), u(q, a) \mid a \in A(q)\}$.

For each random string $r \in R$ of the verifier, we introduce a source-sink pair $s(r), t(r)$, and two collections of edges $E_1(r), E_2(r)$, whose costs are $\infty$. Let $E_1 = \bigcup_{r \in R} E_1(r)$ and $E_2 = \bigcup_{r \in R} E_2(r)$. The set of edges in the final graph is $E_0 \cup E_1 \cup E_2$.

We now fix some random string $r \in R$, and define the set $E_1(r)$ of edges. Let $q = q_1(r)$, $q' = q_2(r)$. Let $(a_1, a_2, \ldots, a_A)$ be any ordering of the set $A(q)$ of answers to $q_1$. For each $1 \leq i \leq A$, let $b_1(a_i), b_2(a_i), \ldots, b_{z_i}(a_i)$ be the set of all answers to $q'$ that are consistent with the answer $a_i$ to $q$. We start by connecting the edges corresponding to $b_1(a_i), b_2(a_i), \ldots, b_{z_i}(a_i)$ into a single path $P_i$ as follows: for $1 \leq j < z_i$, we connect the second endpoint of the edge $e(q', b_j(a_i))$ to the first endpoint of edge $e(q', b_{j+1}(a_i))$. We will refer to $v(q', b_1(a_i))$ as the first vertex on path $P_i$, and to $u(q', b_{z_i}(a_i))$ as the last vertex. Next, we connect the source $s(r)$ to the first vertex of $e(q, a_1)$ and the first vertex of $P_1$. We also connect the second vertex of $e(q, a_A)$ and the last vertex of $P_A$ to the sink $t(r)$. Finally, for all $1 \leq i < A$, we connect the last vertex of $P_i$ to the first vertices of $e(q, a_{i+1})$ and $P_{i+1}$, and the second vertex of $e(q, a_i)$ to the first vertices of $e(q, a_{i+1})$ and $P_{i+1}$. This finishes the definition of the set $E_1(r)$ of edges. Let $G(r)$ be the graph whose vertex set is $V(q) \cup V(q') \cup \{s(r), t(r)\}$, and the edge set consists of $E_1(r)$ and the edges of $E_0$ representing the answers to $q$ and $q'$, that is: $\{e(q, a) \mid a \in A(q)\} \cup \{e(q', a') \mid a' \in A(q')\}$. Then $G(r)$ is an "almost layered" graph, where for each $1 \leq i \leq A$, layer $i$ consists of the edge $e(q_1(r), a_i)$ and of the path $P_i$ (see Figure 7). Notice that the only way to disconnect $s(r)$ from $t(r)$ in graph $G(r)$, without deleting edges of $E_1(r)$ (whose cost is $\infty$), is to delete a pair $e(q, a), e(q', a')$ of edges, where $a$ and $a'$ are matching answers to queries $q$ and $q'$, respectively.

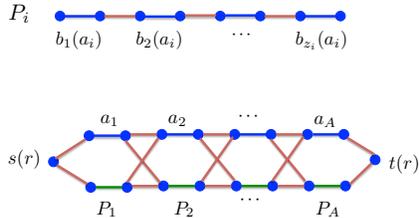

Figure 7: Graph $G(r)$. Red edges belong to $E_1(r)$ and have cost $\infty$.

Finally, we define the sets $E_2(r)$ of edges for all $r \in R$. Given a random string $r \in R$, let $N(r) = N(q_1(r)) \cup N(q_2(r))$, and let $U(r) = \bigcup_{q \in N(r)} V(q)$. Notice that $|U(r)| = |N(q_1(r))| \cdot 7^\ell +$ $|N(q_2(r))| \cdot 2^\ell = 35^\ell + 6^\ell$. We connect $s(r)$ to every vertex in $U(r)$, and we connect every vertex in $U(r)$ to $t(r)$. We denote the resulting set of edges by $E_2(r)$, and we set the costs of these edges to be $\infty$. Finally, we set the parameter $k$ to be $|U(r)| + 1 = 35^\ell + 6^\ell + 1$ (this value is identical for all $r \in R$). Let $G$ be the final instance of the VC-kRC problem.

We now analyze its properties.

**Completeness.** Assume first that $\varphi$ is a Yes-Instance, and consider the strategy of the provers that makes the verifier accept with probability 1. For each query $q \in Q$, let $a(q)$ be the answer to query $q$ under this strategy. Let $E' = \{e(q, a(q)) \mid q \in Q\}$. Notice that the cost of $E'$ is $(5/3)^\ell \cdot |Q_1| + |Q_2| = 2 \cdot (5n/3)^\ell$. Denote this cost by $C_{YI}$. We claim that $E'$ is a feasible solution to the VC-kRC instance. Indeed, consider any random string $r \in R$, and the corresponding source-sink pair $s(r), t(r)$. Let $q = q_1(r), q' = q_2(r)$. Since $a(q), a(q')$ are matching answers to queries $q$ and $q'$, respectively, source $s(r)$ is completely disconnected from sink $t(r)$ in graph $G(r) \setminus E'$. Therefore, any path connecting $s(r)$ to $t(r)$ must use vertices in set $U(r)$. But there are only $|U(r)| = k - 1$ such vertices, so there can be at most $k - 1$ vertex-disjoint paths connecting $s(r)$ to $t(r)$ in $G$.

**Soundness.** Assume now that $\varphi$ is a No-Instance, and let $E'$ be any solution to the VC-kRC instance $G$. We claim that the cost of $E'$ must be at least $C_{YI} \cdot 2^{\gamma\ell/2}/16 = (5n/3)^\ell \cdot 2^{\gamma\ell/2}/8$. Assume otherwise. For each query $q \in Q$, let $A^*(q) = \{a \in A(q) \mid e(q, a) \in E'\}$. We say that a query $q \in Q$ is bad iff $|A^*| \geq 2^{\gamma\ell/2}/2$.

Let $Q'_1$ be the set of all bad queries in $Q_1$ and $Q'_2$ the set of all bad queries in $Q_2$. Notice that $|Q'_1| < |Q_1|/4$: otherwise, $w(E') \geq (5/3)^\ell \cdot 2^{\gamma\ell/2} \cdot |Q'_1|/2 = (5n/3)^\ell \cdot 2^{\gamma\ell/2}/8$, a contradiction. Similarly, $|Q'_2| < |Q_2|/4$. Let $R' \subseteq R$ be the subset of random strings $r$ for which $q_1(r) \notin Q'_1$ and $q_2(r) \notin Q'_2$. Then $|R'| > |R|/2$ must hold, since when we choose a random string $r \in R$ uniformly, the probability that $q_1(r) \in Q'_1$ is less than $1/4$, and the probability that $q_2(r) \in Q'_2$ is less than $1/4$. We now define a strategy for the provers that forces the verifier to accept with probability greater than $2^{-\gamma\ell}$, leading to a contradiction. Given a query $q \in Q_1$, if $q \in Q'_1$, the first prover randomly chooses an answer from the set $A^*(q)$. If $q \notin Q'_1$, the prover returns an arbitrary answer to $q$. Given a query $q \in Q_2$, the strategy of the second prover is defined similarly. We now argue that under this strategy of the two provers, the probability of acceptance is greater than $2^{-\gamma\ell}$. Indeed, a probability that a random string $r \in R'$ is chosen is at least $\frac{1}{2}$. Observe that for each $r \in R'$, there are at most $k - 1 = |U(r)|$ vertex-disjoint paths connecting $s(r)$ to $t(r)$ in graph $G \setminus E'$. However, since the edges of $E_3(r)$ have infinite cost, they do not belong to $E'$. So the pair $s(r)$-$t(r)$ must be completely disconnected in graph $G(r)$: otherwise, if $G(r)$ contains a path $P$ connecting $s(r)$ to $t(r)$, then, together with the edges in $E_3(r)$, we will obtain $k$ vertex-disjoint paths connecting $s(r)$ to $t(r)$. Therefore, there must be an answer $a \in A^*(q)$ to $q$, and an answer $a' \in A^*(q')$, that match. The probability that the first prover selects $a$ and the second prover selects $a'$ is at least $4/2^{\gamma\ell}$. Therefore, overall, the verifier accepts with probability greater than $2^{-\gamma\ell}$, a contradiction.

To summarize, we obtain a gap of $2^{\gamma\ell/2}/16$ between the Yes and the No instances, while the parameter $k = 2^{O(\ell)}$, and the graph size is $n' = n^{O(\ell)}$. Therefore, there is a constant $k_0$, such that for any constant $k > k_0$, we obtain a $k^\epsilon$-hardness of approximation for some specific constant $\epsilon$, if P$\neq$NP. In general, by setting $\ell = \operatorname{poly} \log n$, we get that for any constant $\eta$, for any $k = O\left(2^{(\log n')^{1-\eta}}\right)$, $k > k_0$, there is no $k^\epsilon$-approximation algorithm for VC-kRC unless NP $\subseteq$ DTIME$(n^{\operatorname{poly} \log n})$.

# 8 Single Source-Sink Pair

In this section we study the single source-sink pair version of EC-kRC and VC-kRC, denoted by (st)–EC-kRC and (st)–VC-kRC, respectively. We start with algorithmic results in Section 8.1, and complement them with inapproximability results in Section 8.2.

## 8.1 Algorithms for the Single $(s, t)$-pair Version

This section is devoted to proving Theorem 1.6. Since we show in Section A that VC-kRC captures EC-kRC as a special case, and this reduction remains valid for the single source-sink pair version, it is enough to prove the theorem for VC-kRC. We start with describing the bi-critera approximation algorithm. We reduce the (st)-VC-kRC problem to the problem of finding a minimum-weight vertex (st)-cut in a new graph $G'$. Recall that in this problem, we are given a graph $G'$, with non-negative weights $w(v)$ on vertices $v \in V(G')$, and two special vertices $s$ and $t$. The goal is to find a minimum-weight subset $S \subseteq V(G') \setminus \{s, t\}$ of vertices, whose removal disconnects $s$ from $t$ in $G'$. This problem can be solved efficiently by standard techniques.

Given an instance $G$ of the (st)-VC-kRC problem, let OPT denote the value of the optimal solution (that we guess). We produce an instance $G'$ of the minimum-weight vertex (st)-cut problem, as follows. Graph $G'$ is obtained from graph $G$, after sub-dividing every edge $e \in E(G)$ by a vertex $v_e$. The weight of this new vertex is set to be $w(v_e) = w_e$, and for each original vertex $v \in V(G)$, we set its weight to $\frac{c}{k-1} \cdot \mathsf{OPT}$.

Assume that we have guessed the value OPT correctly, and let $E'$ be the optimal solution to the (st)-VC-kRC problem on graph $G$, with $w(E') = \mathsf{OPT}$. Then graph $G \setminus E'$ contains a subset $S' \subseteq V(G) \setminus \{s, t\}$ of at most $(k - 1)$ vertices, whose removal disconnects $s$ from $t$. Let $S'' = \{v_e \mid e \in E'\}$ be the subset of vertices of $G'$ corresponding to the edges in $E'$. Then $S' \cup S''$ is a feasible solution to the vertex (st)-cut problem in graph $G'$, and its value is $w(S') + w(S'') \leq (k-1) \cdot \frac{c}{k-1} \cdot \mathsf{OPT} + \mathsf{OPT} \leq (1+c) \cdot \mathsf{OPT}$. Therefore, the value of the minimum-weight (st)-cut in $G'$ is at most $(1 + c) \cdot \mathsf{OPT}$. On the other hand, let $S$ be the minimum-weight vertex (st)-cut in graph $G'$. Partition $S$ into two subsets: $S' = S \cap V(G)$ is the subset of vertices in the original graph $G$, and $S'' = S \setminus S'$ is the set of all remaining vertices. Let $E' = \{e \mid v_e \in S''\}$ be the corresponding subset of edges in graph $G$. Then $|S'| \leq \frac{\mathsf{OPT} \cdot (1+c)}{c \cdot \mathsf{OPT}/(k-1)} \leq (k-1)(1 + 1/c) < k(1 + 1/c)$. Therefore, $E'$ is a $((1 + 1/c), (1 + c))$-bi-criteria approximate solution.

In order to obtain a factor $k$-approximation algorithm, we use the above algorithm, setting the parameter $c = k$. Using the above analysis, the value of the minimum-weight node (st)–cut in graph $G'$ is at most $(k + 1) \cdot \mathsf{OPT}$. Moreover, since $|S'| \leq (k-1)(1 + 1/k) = k - 1 + \frac{k-1}{k} < k$, set $E'$ of edges is indeed a feasible solution for the VC-kRC instance $G$.

## 8.2 Inapproximability of (st)–VC-kRC

In this section we complement our upper bounds from Section 8.1 by inapproximability results, and prove Theorems 1.7, 1.8 and 1.9. The starting point for all three reductions is similar. We define the Small Set Vertex Expansion (SSVE) problem, and show an approximation preserving reduction from SSVE to (st)–VC-kRC, in Section 8.2.1. We then show inapproximability results for SSVE in subsequent sections, which are used to establish the lower bounds on the approximability of (st)–VC-kRC.

### 8.2.1 Small Set Vertex Expansion

**Definition 8.1** (SMALL SET VERTEX EXPANSION PROBLEM (SSVE)). *Given a bipartite graph $G = (U, V, E)$ and a parameter $0 < \alpha \leq 1$, the aim is to find a subset $S \subseteq U$ of vertices, $|S| \geq \alpha |U|$, minimizing the number of its neighbors, $|\Gamma(S)|$.*

We present a gap-preserving reduction from SSVE to (st)-VC-kRC, that will allow us to later focus on proving inapproximability of SSVE.

**Theorem 8.2** *Let $G = (U, V, E)$ be any bipartite graph with $|U| = m, |V| = n$, and let $N = 2mn + 1$. We can efficiently construct an edge-weighted graph $G'$ with two special vertices $s, t \in V(G')$, such that for any $0 < \alpha < 1$, and any integer $0 \leq C \leq |V|$, the following property holds: there is a subset $S \subseteq U$ in graph $G$ with $|S| \geq \alpha |U|$ and $|\Gamma(S)| \leq C$ iff there is a solution of cost at most $C \cdot N$ to the (st)-VC-kRC problem on graph $G'$, where the parameter $k$ is set to be $k = |U|(1 - \alpha) + 1$.*

**Proof:** Given an SSVE instance $G = (U, V, E)$, with $|U| = m, |V| = n$, let $N = 2nm + 1$. In order to construct the graph $G'$, we start with the bipartite graph $G = (U, V, E)$, and then replace every vertex $v \in V$ with a clique $K(v)$ on $N$ new vertices. All edges of the clique $K(v)$ have cost $\infty$. Let $V' = \bigcup_{v \in V} V(K(v))$ be the set of all vertices in all such cliques. We add an edge of cost $\infty$ between $u \in U$ and every vertex in $K(v)$ if $(u, v) \in E(G)$. We also add two additional vertices $s$ and $t$. For every vertex $u \in U$, add an $\infty$-cost edge $(s, u)$, and for every $v' \in V'$, add a cost-1 edge $(v', t)$ to $G'$. This completes the description of graph $G'$. Given a parameter $0 < \alpha < 1$, we set $k = |U|(1 - \alpha) + 1$.

*Completeness.* Suppose we have a subset $S \subseteq U$ with $|S| = \alpha |U|$ and $|\Gamma(S)| \leq C$ in graph $G$. We construct a solution to the (st)-VC-kRC instance $G'$, as follows: for each vertex $v \in \Gamma(S)$, we add all edges between the vertices of $K(v)$ and the vertex $t$ to the solution. Let $E^*$ denote the resulting set of edges. Then $|E^*| \leq CN$. We now argue that $E^*$ is a valid $k$-route (st)-cut. Indeed, consider the graph $G' \setminus E^*$, and let $S' = U \setminus S$. Then $|S'| = k - 1$, and once the vertices of $S'$ are removed from $G'$, no paths connecting $s$ to $t$ remain in the graph.

*Soundness.* Assume now that we have a solution $E^*$ of cost at most $CN$ to the (st)-VC-kRC instance $G'$. Notice that all edges in $E^*$ must be incident on $t$, since all other edges have cost $\infty$.

Our first step is to transform the solution $E^*$, so that for each vertex $v \in V$, either all edges connecting the vertices of $K_v$ to $t$ belong to the solution, or none of them. In order to perform this transformation, we consider the vertices $v \in V$ one-by-one. For each such vertex $v$, let $E_v$ be the set of all edges connecting the vertices of $K(v)$ to $t$. If $|E_v \setminus E^*| \geq k$, then we remove all edges of $E_v$ from $E^*$. Otherwise, we add all edges of $E_v \setminus E^*$ to $E^*$. We first claim that the resulting set of edges remains a valid solution to the (st)-VC-kRC problem. Indeed, let $E^*$ be the subset of edges in the solution before the vertex $v \in V$ is processed, and assume that $E^*$ is a valid $k$-route cut. Assume that $|E_v \setminus E^*| \geq k$. Partition the vertices in $K(v)$ into two subsets: $V_1$ is the subset of vertices $v'$ whose edge $(v', t) \in E^*$, and $V_2$ denotes the set of the remaining vertices. Notice that $|V_2| \geq k$ must hold. We claim that $E^* \setminus E_v$ remains a valid solution to the $k$-route cut instance. Assume otherwise. Then graph $G' \setminus (E^* \setminus E_v)$ has $k$ vertex-disjoint paths $P_1, \ldots, P_k$, connecting $s$ to $t$. We can assume w.l.o.g. that each such path contains at most one vertex of $K(v)$, since each vertex in $K(v)$ is directly connected to $t$ by an edge in graph $G' \setminus (E^* \setminus E_v)$. For $1 \leq i \leq k$, let $v_i \in K(v)$ be the vertex lying on $P_i$ (if such exists). We now construct $k$ vertex-disjoint paths $P'_1, \ldots, P'_k$, connecting $s$ to $t$ in graph $G' \setminus E^*$, reaching a contradiction. Each path $P'_i$ is constructed from path $P_i$, by replacing the vertex $v_i$ with some vertex of $V_2$, so that each vertex of $V_2$ appears on at most one such path. This can be done since $|V_2| \geq k$.

Let $E^{**}$ be the subset of edges obtained from $E^*$ after we process all vertices $v \in V$. From the above discussion, $E^{**}$ is a feasible solution to the (st)-VC-kRC instance $G'$. Moreover, $|E^{**} \setminus E^*| \leq nk \leq 2mn < N$, and $|E^{**}|$ is an integral multiple of $N$. Therefore, if $|E^*| \leq CN$, where $C$ is an integer, $|E^{**}| \leq CN$ as well.

Since $E^{**}$ is a valid $k$-route cut in $G'$, there is a set $S'$ of $(k-1)$ vertices, whose removal from $G' \setminus E^{**}$ separates $s$ from $t$. Since for each $v \in V$, either $E_v \subseteq E^{**}$, or $E_v \cap E^{**} = \emptyset$, while $|K(v)| > 2k$, we can assume w.l.o.g. that $S' \subseteq U$. Let $S = U \setminus S'$. Then $|S| \geq \alpha m$, and the set $\Gamma(S)$ is contained in the set $\{v \in V \mid E_v \subseteq E^{**}\}$, implying that $|\Gamma(S)| \leq C$. $\square$

### 8.2.2 Inapproximability from the Random $\kappa$-AND Assumption

This section is devoted to proving Theorem 1.7. We prove the following inapproximability result for SSVE.

**Theorem 8.3** *For every large enough constant $\kappa$, there are parameters $\alpha, \beta$ that depend on $\kappa$ only, such that, assuming Hypothesis 1.1, no polynomial-time algorithm, given a bipartite graph $G = (U, V, E)$, can distinguish between the following two cases:*

- *Completeness: there is a subset $S \subseteq U$, with $|S| = \alpha|U|$ and $|\Gamma(S)| \leq \beta|V|$.*

- *Soundness: for any subset $S \subseteq U$ with $|S| \geq \frac{\alpha}{2^{\kappa/2}}|U|$, $|\Gamma(S)| > \beta \cdot 2^{\sqrt{\kappa}/c}|V|$,*

*where $c$ is a constant independent of $\kappa$.*

Combining Theorem 8.3 with Theorem 8.2, we get that there is no polynomial-time algorithm for (st)–VC-kRC, that distinguishes between the cases where there is a solution of cost $\beta|V| \cdot N$ for parameter $k = |U|(1-\alpha)+1$, and the cases where there is no solution of cost $\beta/2^{\sqrt{\kappa}/c}$ and parameter $k = |U|\left(1 - \frac{\alpha}{2^{\kappa/2}}\right) + 1$. Since $\alpha$ and $\kappa$ are constants, this will complete the proof of Theorem 1.7.
**Proof:** The proof proceeds in two steps. In the first step, we show a simple reduction that gives a weak inapproximability result for the SSVE problem. Next, we amplify the inapproximability factor, by using graph products. The first step is summarized in the next lemma.

**Lemma 8.4** *Let $0 < \delta < \frac{1}{2}$ be any parameter, and $\kappa > \kappa_0$ any large enough constant, where $\kappa_0$ depends on $\delta$. Assuming Hypothesis 1.1, there is no polynomial time algorithm, that, given a bipartite graph $G = (U, V, E)$, can distinguish between the following two cases:*

- *Completeness: There is a subset $S \subseteq U$, with $|S| = |U|/2^{c_0\sqrt{\kappa}}$ and $|\Gamma(S)| \leq |V|/2$.*

- *Soundness: For any subset $S \subseteq U$, with $|S| \geq |U|/2^{\kappa(1-2\delta)}$, $|\Gamma(S)| > (1+\delta) \cdot |V|/2$.*

*(here $c_0$ is the constant from Hypothesis 1.1).*

**Proof:** We start with the following simple claim about random instances of the $\kappa$-AND problem.

**Claim 8.1** *Let $0 < \delta < \frac{1}{2}$ be any parameter, $\kappa > \kappa_0$ a large enough constant, where $\kappa_0$ depends on $\delta$, and $\Delta \geq \Delta_0$ a large enough constant, where $\Delta_0$ depends on $\kappa$ and $\delta$. Then for any random $\kappa$-AND formula $\Phi$ on $n$ variables and $m = \Delta n$ clauses, every subset of $m \cdot 2^{2\delta\kappa}/2^\kappa$ clauses in $\Phi$ contains at least $(1+\delta)n$ different literals, with high probability.*

**Proof:** Fix any set $S$ of $(1+\delta)n$ different literals. We say that a clause $C$ of $\Phi$ is bad for $S$, iff all literals of $C$ belong to $S$. Let $\mathcal{E}(C, S)$ denote the event that $C$ is bad for $S$. Then:

$$\mathbf{Pr}\left[\mathcal{E}(C, S)\right] = \left(\frac{1+\delta}{2}\right)^\kappa < 2^{1.5\delta\kappa}/2^\kappa$$

Therefore, the expected number of clauses $C \in \Phi$ that are bad for $S$ is at most $\mu = m \cdot 2^{1.5\delta\kappa}/2^\kappa$. We say that a bad event $\mathcal{E}(S)$ happens if at least $m \cdot 2^{2\delta\kappa}/2^\kappa$ clauses of $\Phi$ have their literals contained in $S$. Notice that $m \cdot 2^{2\delta\kappa}/2^\kappa = \mu \cdot 2^{0.5\delta\kappa} = \mu(1+\delta')$ for some constant $\delta'$ that depends on $\delta$ and $\kappa$. Therefore, by Chernoff bounds,

$$\mathbf{Pr}\left[\mathcal{E}(S)\right] \leq e^{-\mu\delta'^2/2}$$

Let $\mathcal{E}$ be the event that for any subset $S$ of $(1+\delta)n$ literals, the bad event $\mathcal{E}(S)$ happens. Since the total number of subsets $S$ of literals is at most $2^{2n}$, using the union bound,

$$\mathbf{Pr}\left[\mathcal{E}\right] \leq 2^{2n} \cdot e^{-\mu\delta'^2/2} = 2^{2n} \cdot 2^{-c' \cdot m}$$

where $c'$ is some constant that depends on $\delta$ and $\kappa$. Clearly, setting $\Delta = 4/c'$ ensures that $\mathbf{Pr}\left[\mathcal{E}\right] \leq 2^{-2n}$. □

Given a Random $\kappa$-AND instance $\Phi$ with $n$ variables and $m = \Delta n$ variables (where $\Delta$ is chosen as in Claim 8.1), we define the corresponding bipartite graph $G = (U, V, E)$ as follows. Let $U$ be the set of clauses, and $V$ be the set of $2n$ literals. Connect each clause $C$ and literal $\ell$ with an edge iff $\ell$ belongs to $C$.

For the completeness case, if $\Phi$ is $(2^{-c_0\sqrt{\kappa}})$-satisfiable, let $S$ be the set of $m/2^{c_0\sqrt{\kappa}}$ clauses that can be satisfied by some assignment. It is easy to see that $\Gamma(S)$ never contains both a literal and its negation. Therefore, $|\Gamma(S)| \leq n = |V|/2$.

For the soundness case, by Claim 8.1, with high probability over the choice of the $\kappa$-AND formula $\Phi$, for every set $S \subseteq V$ of size $|S| = m/2^{\kappa(1-2\delta)}$, we have $|\Gamma(S)| > (1+\delta) \cdot n = (1+\delta)|V|/2$. Therefore, an efficient algorithm that distinguishes between the completeness and the soundness settings can be used to refute random $\kappa$-AND formulas. □

We say that a bipartite graph $G = (U, V, E)$ is $(\alpha, \beta)$-expanding iff for every $S \subseteq U$, with $|S| \geq \alpha|U|$, we have $|\Gamma(S)| > \beta|V|$. From Lemma 8.4, assuming Hypothesis 1.1, no polynomial-time algorithm can distinguish between bipartite graphs that are $\left(2^{-\kappa(1-2\delta)}, \frac{1+\delta}{2}\right)$-expanding, and graphs that are not $\left(2^{-c_0\sqrt{\kappa}}, \frac{1}{2}\right)$-expanding, for any constant $0 < \delta < \frac{1}{2}$ and any large enough constant $\kappa > \kappa_0$, where $\kappa_0$ depends on $\delta$. Next, we will use tensor product of bipartite graphs to amplify this gap.

**Definition 8.5** (TENSOR PRODUCT OF BIPARTITE GRAPHS)
For two graphs $G_1 = (U_1, V_1, E_1)$, $G_2 = (U_2, V_2, E_2)$, let $G = G_1 \otimes G_2$ be the bipartite graph $G = (U, V, E)$, where $U = U_1 \times U_2$, $V = V_1 \times V_2$, and

$$E = \{((u_1, u_2), (v_1, v_2)) \mid (u_1, u_2) \in U_1 \times U_2, (v_1, v_2) \in V_1 \times V_2, (u_1, v_1) \in E_1, (u_2, v_2) \in E_2\}$$

For $k \in \mathbb{Z}^+$, define $G^{\otimes k}$ inductively as follows: $G^{\otimes 2} = G \otimes G$, and for $k > 2$, $G^{\otimes k} = G^{\otimes(k-1)} \otimes G$.

The following lemma shows that the tensor product can be used to amplify the expansion gap, with a small loss in the threshold parameter $\alpha$.

**Lemma 8.6** *[Tensor product amplification]* Let $G = (U, V, E)$ be any bipartite graph, and $0 < \alpha, \beta < 1$ parameters.

- If $G$ is not $(\alpha, \beta)$-expanding, then $G^{\otimes 2}$ is not $(\alpha^2, \beta^2)$-expanding.

- If $G$ is $(\alpha, \beta)$-expanding, then $G^{\otimes 2}$ is $(2\alpha - \alpha^2, \beta^2)$-expanding, and therefore is $(2\alpha, \beta^2)$-expanding.

**Proof:** Let $G^{\otimes 2} = (U', V', E')$, and let $m = |U|$, $n = |V|$. Then $|U'| = m^2, |V'| = n^2$. Assume first that $G$ is not $(\alpha, \beta)$-expanding. Then there is a subset $S \subseteq U$ of vertices, $|S| = \alpha m$, such that $|\Gamma_G(S)| \leq \beta n$. Consider the subset $S \times S$ of vertices in graph $G^{\oplus 2}$. Then $|S \times S| = \alpha^2 m^2$, and $|\Gamma_{G^{\otimes 2}}(S \times S)| = |\Gamma(S) \times \Gamma(S)| \leq \beta^2 n^2$. Therefore, graph $G^{\otimes 2}$ is not $(\alpha^2, \beta^2)$-expanding.

Assume now that $G$ is $(\alpha, \beta)$-expanding, and assume for contradiction that $G^{\otimes 2}$ is not $(2\alpha - \alpha^2, \beta^2)$-expanding. Let $S' \subseteq U \times U$ be a subset of $V'$ with $|S'| = (2\alpha - \alpha^2)m^2$, and $|\Gamma_{G^{\otimes 2}}(S')| \leq \beta^2 n^2$. For each $i \in U$, let $S'_i = \{j \mid (i, j) \in S'\}$. Call $i \in U$ *good* iff $|S'_i| \geq \alpha |U|$. Let $T \subseteq U$ be the set of all good vertices. By an averaging argument, we get that

$$|S'| \leq |T| \cdot m + (m - |T|) \cdot \alpha m$$

Therefore, $|T| \geq \frac{|S'| - \alpha m^2}{(1-\alpha)m} \geq \frac{(2\alpha - \alpha^2)m^2 - \alpha m^2}{(1-\alpha)m} = \alpha m$.

On the other hand, for each $i \in T$, the number of neighbors of the vertices in $\{i\} \oplus S'_i$ in graph $G^{\oplus 2}$ is at least $|\Gamma_G(i)| \cdot |\Gamma_G(S_i)| \geq |\Gamma_G(i)| \cdot \beta n$ (since $|S_i| \geq \alpha m$, and graph $G$ is $(\alpha, \beta)$-expanding). Therefore, $|\Gamma_{G^{\otimes 2}}(S')| \geq |\Gamma_G(T)| \cdot \beta n > \beta^2 n^2$, a contradiction to $S'$ being a violating set. □

We are now ready to complete the proof of Theorem 8.3. We start with the instances given by Lemma 8.4, and repeatedly apply Lemma 8.6 to them. Specifically, let $G_0$ be the graph obtained from Lemma 8.4, and for $i > 0$, let $G_i = G_{i-1}^{\otimes 2}$. Our final graph is $G_\ell$, where $\ell = \log(\sqrt{\kappa}/6c_0)$.

Assume first that the initial graph $G_0$ is a YES-instance, that is, $G_0$ is not $(\alpha_0, \beta_0)$-expanding, for $\alpha_0 = 2^{-c_0\sqrt{\kappa}}$, $\beta_0 = \frac{1}{2}$. Then by Lemma 8.6, graph $G_\ell$ is not $(\alpha_\ell, \beta_\ell)$-expanding, where $\alpha_\ell = \alpha_0^{2^\ell} = 2^{-c_0\sqrt{\kappa} \cdot \sqrt{\kappa}/6c_0} = 2^{-\kappa/6}$, and $\beta_\ell = \beta_0^{2^\ell} = 2^{-\sqrt{\kappa}/6c_0}$.

Assume now that the initial graph $G_0$ is a NO-instance, that is, $G_0$ is $(\alpha_0, \beta_0)$-expanding, for $\alpha_0 = 2^{-k(1-2\delta)}$, $\beta_0 = (1+\delta)/2$. Then by Lemma 8.6, graph $G_\ell$ is $(\alpha_\ell, \beta_\ell)$-expanding, for $\alpha_\ell = 2^\ell \alpha_0 = \frac{\sqrt{\kappa}}{6c_0 \cdot 2^{k(1-2\delta)}}$, and $\beta_\ell = \beta_0^{2^\ell} = \left(\frac{1+\delta}{2}\right)^{\sqrt{\kappa}/6c_0}$.

Let us now set $\delta = 1/18$, and denote $\alpha = 2^{-\kappa/6}$ and $\beta = 2^{-\sqrt{\kappa}/6c_0}$. Denote $G_\ell = (U, V, E)$, with $|U| = m$, $|V| = n$. We then get that in the YES-instance, there is a subset $S \subseteq U$ of at least $\alpha|U|$ vertices, such that $|\Gamma(S)| \leq \beta|V|$. For the No-instance, for any subset $S \subseteq U$ with $|S| \geq \alpha|U|/g_1$, we have that $|\Gamma(S)| > \beta|V| \cdot g_2$. It now remains to bound $g_1$ and $g_2$. First,

$$g_1 = \frac{1}{2^{\kappa/6}} / \frac{\sqrt{\kappa}}{6c_0 \cdot 2^{\kappa(1-2\delta)}} \geq \frac{2^{\kappa(1-3\delta)}}{2^{\kappa/6}} \geq 2^{\kappa/2}$$

if $\kappa$ is large enough and $\delta = 1/18$.

Finally,

$$g_2 = \left(\frac{1+\delta}{2}\right)^{\sqrt{\kappa}/6c_0} / 2^{-\sqrt{\kappa}/6c_0} = (1+\delta)^{\sqrt{\kappa}/6c_0}$$

This completes the proof of the bicriteria hardness in Theorem 8.3. □

### 8.2.3 Inapproximability from the Random 3SAT Assumption

In this section we focus on proving Theorem 1.8. We use the following re-statement of Theorem 2 of Feige [Fei02].

**Theorem 8.7** *For every fixed $\epsilon > 0$, for a $\Delta$ sufficiently large constant independent of $n$, assuming Hypothesis 1.2, there is no polynomial-time algorithm, that, given a random 3AND formula on $n$ variables and $m = \Delta n$ clauses, returns "typical" with probability $\frac{1}{2}$, but never returns "typical" if the formula is $\left(\frac{1}{4} - \epsilon\right)$-satisfiable.*

As before, we start by proving a bi-criteria hardness result for the SSVE problem.

**Theorem 8.8** *Given a bipartite graph $G = (U, V, E)$, assuming Hypothesis 1.2, no polynomial time algorithm can distinguish between the following two cases:*

- *Completeness: there is a subset $S \subseteq U$ with $|S| \geq |U|/5$ and $|\Gamma(S)| \leq |V|/2$, and*

- *Soundness: for every subset $S \subseteq U$, if $|S| \geq |U|/6$, then $|\Gamma(S)| \geq 11|V|/20$.*

Theorem 1.8 then follows immediately by combining Theorem 8.2 with Theorem 8.8. We now focus on proving Theorem 8.8. As before, we first prove a simple fact about random 3AND formulas, that will lead to the proof of the theorem.

**Claim 8.2** *For sufficiently large $\Delta$, with high probability, every set of $m/6$ clauses in an R3AND instance with $m = \Delta n$ clauses contains at least $1.1n$ different literals.*

**Proof:** Fix a set $S$ of $1.1n$ different literals. For a 3AND clause $C$, let $\mathcal{E}(S, C)$ be the event that all three literals of $C$ are contained in $S$. Clearly, $\mathbf{Pr}\left[\mathcal{E}(S, C)\right] = \left(\frac{1.1}{2}\right)^3 \leq 1/6 - c$ for some constant $c > 0$. Therefore, the expected number of clauses contained in $S$ is at most $(1/6 - c)m$. Let $\mathcal{E}(S)$ be the bad event that at least $m/6$ clauses are contained in $S$.

By Chernoff bound,

$$\Pr[\mathcal{E}(S)] < 2^{-c'm},$$

for some small constant $c' > 0$.

Let $\mathcal{E}$ be the event that $\mathcal{E}(S)$ happens for at least one subset $S$ of $1.1n$ literals. Since the number of such possible subsets $S$ is bounded by $2^{2n}$, when $\Delta > 3/c'$, by a union bound, we have that

$$\Pr[\mathcal{E}] < 2^{-c'm} \cdot 2^{2n} < 2^{-n}.$$

□

Given a random 3AND formula $\Phi$, we construct a graph $G = (U, V, E)$, where $U$ contains a vertex $u_C$ for each clause $C \in \Phi$, and $V$ contains a vertex $v_\ell$ for each literal $\ell$. We add an edge $(u_C, v_\ell)$ iff literal $\ell$ belongs to clause $C$. Assume for contradiction that there is a polynomial-time algorithm $\mathcal{A}$, distinguishing between instances where there is a subset $S \subseteq U$ with $|S| \geq |U|/5$ and $|\Gamma(S)| \leq |V|/2$, and instances where for every subset $S \subseteq U$, if $|S| \geq |U|/6$, then $|\Gamma(S)| \geq 11|V|/20$. Given a random 3AND formula $\Phi$, we apply algorithm $\mathcal{A}$ to the resulting graph $G$. If the algorithm establishes that we are in the second scenario (that is, for each subset $S$ with $|S| \geq |U|/6$, $\Gamma(S) \geq 11|V|/20$), then we output "typical". From Claim 8.2, this will happen most of the time. However, if the formula $\Phi$ is $\left(\frac{1}{4} - \epsilon\right)$-satisfiable, then we can let $S$ be the set of satisfied clauses, and since $\Gamma(S)$ cannot contain a literal and its negation, we will get that $|\Gamma(S)| \leq |V|/2$. Therefore, using Theorem 8.7, algorithm $\mathcal{A}$ can be used to refute Hypothesis 1.2.

### 8.2.4 Reduction from the Densest $\kappa$-Subgraph Problem

In this section we prove Theorem 1.9. As before, we do so by proving a similar result for the SSVE problem.

**Theorem 8.9** *For any constant $\lambda \geq 2$, and for any approximation factor $\rho$ (that may depend on $n$), if there is an efficient factor $\rho$ approximation algorithm for the SSVE problem, then there is an efficient factor $(2\rho^\lambda)$-approximation algorithm for the $\lambda$-uniform Hypergraph Densest $\kappa$-subgraph problem.*

Observe that combining Theorem 8.9 with Theorem 8.2 immediately implies Theorem 1.9. We now focus on proving Theorem 8.9.

**Proof:** Given a $\lambda$-uniform Hypergraph Densest $\kappa$-subgraph instance $G = (V, E)$, we construct an instance $G' = (U', V', E')$ of SSVE as follows. For each hyper-edge $e \in E$, we add a vertex $u_e$ to $U'$. The set $V'$ of vertices is $V' = V$. We add an edge between $u_e \in U'$ and $v \in V'$ iff vertex $v$ belongs to the hyper-edge $e$. Since $\lambda$ is a constant, we can assume that $\kappa \gg \lambda^2$ (otherwise, the optimal solution to the Densest $\kappa$-subgraph instance can be found efficiently by exhaustive search).

Let $\mathcal{A}$ be a factor $\rho$ approximation algorithm for the SSVE problem. We now show a factor $2\rho^\lambda$-approximation algorithm for the Densest $\kappa$-subgraph problem. The algorithm will guess the value $m'$ of the optimal solution to the Densest $k$-subgraph instance $G$. It will then apply algorithm $\mathcal{A}$ to instance $G'$, with value $\alpha = m'/|U'|$. If value $m'$ was guessed correctly, then there is a subset $S \subseteq U'$ of vertices, with $|S| = m'$, and $|\Gamma(S)| = \kappa$. Therefore, algorithm $\mathcal{A}$ must return a subset $S' \subseteq U'$ of vertices, with $|S'| = m'$, and $|\Gamma(S')| \leq \kappa \cdot \rho$. Let $V' = \Gamma(S')$. Set $V'$ is also a subset of vertices in the initial instance $G$, and we are now guaranteed that $|V'| \leq \kappa \cdot \rho$, while the number of edges contained in $V'$ is at least $m'$. Let $V''$ be a random subset of $\kappa$ vertices from $V'$. Observe that for a hyper-edge $e \subseteq V'$, the probability that $e$ is contained in $V''$ is at least $\left(\frac{\kappa-\lambda}{\rho\kappa}\right)^\lambda \geq \frac{2}{3} \cdot \left(\frac{1}{\rho}\right)^\lambda$ (since $\kappa \gg \lambda^2$), and so the expected number of hyper-edges contained in $V''$ is at least $2m'/3\rho^\lambda$. □

## Acknowledgements

The first author would like to thank Sanjeev Khanna for suggesting the problem and for many interesting discussions. We would also like to thank Rajsekar Manokaran for helpful discussions on the use of the random $\kappa$-AND conjecture.

## A  Connection between VC-kRC and EC-kRC

In this section we show that EC-kRC can be cast as a special case of VC-kRC. Indeed, assume that we are given an instance $G = (V, E)$ of EC-kRC with costs $w_e$ on edges $e \in E$, a set $\{(s_i, t_i)\}_{i=1}^r$ of source-sink pairs, and an integer $k$. We construct an instance $G' = (V', E')$ of the VC-kRC problem as follows. For each vertex $u \in V$, for each edge $e$ incident on $u$ in $G$, we create a new vertex $v(u, e)$ in graph $G'$. Each pair $v(u, e), v(u, e')$ of such vertices is connected with an edge of cost $\infty$. In other words, we replace the vertex $u$ with a clique of $d(u)$ vertices, where $d(u)$ is the degree of $u$ in $G$, and set the weights of the edges in the clique to be $\infty$. Let $K(u)$ denote this clique, and let $V(u)$ and $E(u)$ denote the sets of its vertices and edges, respectively. Let $E_1 = \bigcup_{u \in V} E(u)$. We now define another set $E_2$ of edges, corresponding to the original edges in graph $G$. For each edge $e = (u, u') \in E$, we add an edge $(v(u, e), v(u', e))$ of weight $w_e$ to $E_2$. Graph $G' = (V', E')$ is then defined as: $V' = \bigcup_{u \in V} V(u)$, and $E' = E_1 \cup E_2$. In order to define the source-sink pairs, we select, for each $1 \leq i \leq r$, an arbitrary vertex $s'_i \in V(s_i)$, and an arbitrary vertex $t'_i \in V(t_i)$, and we let $(s'_i, t'_i)$ be a source-sink pair in the new instance. Therefore, the set of the source-sink pairs becomes $\{(s'_i, t'_i)\}_{i=1}^r$. The parameter $k$ remains unchanged. We now show that the two instances are equivalent, in the sense that any feasible solution $E^*$ to the EC-kRC instance $G$ implies a feasible solution $E^{**}$ to the VC-kRC instance $G'$, and vice versa.

Let $E^*$ be any feasible solution to the EC-kRC instance $G$. We claim that there is a solution $E^{**}$ to the VC-kRC instance $G'$, of the same weight. The solution $E^{**}$ contains, for each edge $e = (u, u') \in E^*$, the corresponding edge $(v(u, e), v(u', e))$ of $E'$. Clearly, the weight of $E^{**}$ is the same as the weight of $E^*$. We now claim that $E^{**}$ is a feasible solution to the VC-kRC instance $G'$. Assume otherwise, and let $(s'_i, t'_i)$ be any source-sink pairs, such that $G' \setminus E^{**}$ contains at least $k$ node-disjoint paths $P'_1, \ldots, P'_k$ connecting $s'_i$ to $t'_i$. We show that graph $G \setminus E^*$ must then contain at least $k$ edge-disjoint paths $P_1, \ldots, P_k$, connecting $s_i$ to $t_i$, leading to a contradiction. For $1 \leq j \leq k$, path $P_j$ is constructed from path $P'_j$, as follows. Let $(s'_i = v_0, v_1, \ldots, v_z = t'_i)$ be the sequence of vertices on path $P'_j$. For each vertex $v_{z'}$ on this path, if $v_{z'} \in K(u_{z'})$, then we replace $v_{z'}$ with $u_{z'}$. Let $P_j$ be the resulting path, after we erase possible cycles and consecutive occurrences of the same vertex on it. Then $P_j$ is a valid $s_i - t_i$ path in graph $G$, and moreover, since paths $P'_1, \ldots, P'_k$ were vertex-disjoint, this ensures that the paths $P_1, \ldots, P_k$ are edge-disjoint.

Assume now that $E^{**}$ is any feasible solution of weight less than $\infty$ to instance $G'$ of VC-kRC. Then all edges in $E^{**}$ belong to set $E_2$. Let $E^*$ be the set of corresponding edges in graph $G$. Clearly, the weight of $E^*$ is the same as the weigh of $E^{**}$. We only need to show that $E^*$ is a feasible solution to the EC-kRC instance $G$. Assume otherwise, and let $(s_i, t_i)$ be a source-sink pair, such that $G \setminus E^*$ contains at least $k$ edge-disjoint paths $P_1, \ldots, P_k$ connecting $s_i$ to $t_i$. We show a collection $P'_1, \ldots, P'_k$ of node-disjoint paths, connecting $s'_i$ to $t'_i$ in $G' \setminus E^{**}$, thus obtaining a contradiction. Fix some $1 \leq j \leq k$. Path $P'_j$ is constructed from path $P_j$ as follows. Let $(s_i = u_0, u_1, \ldots, u_z = t_i)$ be the sequence of vertices on path $P_j$. For each $0 \leq z' < z$, let $e_{z'} = (u_{z'}, u_{z'+1})$. For $1 \leq z' \leq z - 1$, we replace the vertex $u_{z'}$ on the path by two vertices: $v(u_{z'}, e_{z'-1})$ and $v(u_{z'}, e_{z'})$. If $s'_i = v(s_i, e_0)$, then we replace $s_i$ with $s'_i$. Otherwise, we replace it with a pair $s'_i, v(s_i, e_0)$ of vertices. Similarly, if $t'_i = v(t_i, e_{z-1})$, then we replace $t_i$ with $t'_i$. Otherwise, we replace it with $v(t_i, e_{z-}), t'_i$. Let $P'_j$ denote the resulting path. It is easy to see that this is a valid path in graph $G'$. Moreover, if paths $P_1, \ldots, P_k$ were edge-disjoint, the paths $P'_1, \ldots, P'_k$ are guaranteed to be vertex-disjoint in graph $G'$.

## B  VC-kRC: from General to Uniform Edge Costs

In this section we show that in the VC-kRC problem, we can assume w.l.o.g. that all edge weights are unit. We will lose an $(1 + 1/n)$ factor in the approximation ratio in this transformation.

Consider any VC-kRC instance $G(V, E)$ where $w : E \to \Re^+$ are non-negative edge weights. Let $w_{max}, w_{min}$ be the maximum and minimum weights across all edges. To obtain an unweighted version, let us first consider the case when $w_{max}/w_{min} < n^4$. We can then assume w.l.o.g. that $w_{min} = 1$ and $w_{max} < n^4$. Round the weight of every edge up to the next multiple of $1/n^3$, and multiply all edge weights by $n^3$, obtaining integral weights $w'_e$. We now replace every edge $e$ by $w'(e)$ parallel edges. The crucial observation is that two parallel edges can not be on different node disjoint paths, and so, in the optimal solution for the new instance, for each original edge $e \in E$, either all its copies belong to the solution, or none of them. The cost of the optimal solution in the new instance increases by the additive factor of $1/n$ due to the rounding of the edge weights, and since we have assumed that all edge weights are at least 1, we lose at most $(1 + 1/n)$-factor in the solution cost.

When $w_{max}/w_{min}$ is not bounded by $n^4$, we first guess the value of OPT of the optimal solution, and delete all edges $e$ with $w(e) < \frac{\text{OPT}}{n^3}$. These edges will be eventually added to our solution. Note that all such edges can contribute at most $\text{OPT}/n$ to the solution value. For all edges of weight more than $n \cdot \text{OPT}$, we set their new weight to be $n \cdot \text{OPT}$, and we repeat the reduction mentioned above. It is easy to see the the same argument works here as well.